\documentclass[5p]{elsarticle}
\makeatletter
\def\ps@pprintTitle{%
 \let\@oddhead\@empty
 \let\@evenhead\@empty
 \def\@oddfoot{}%
 \let\@evenfoot\@oddfoot}
\makeatother
\usepackage{hyperref}
\usepackage{amsmath}
\usepackage{amsfonts}
\usepackage{amssymb}
\usepackage{enumitem}
\usepackage{physics}
\usepackage{bbm}
\usepackage{stmaryrd}
\usepackage{extarrows}
\usepackage{color}
\usepackage{amsthm}

\usepackage{bm}
\usepackage{wrapfig}

\usepackage{mathrsfs}
\usepackage{setspace}
\usepackage{natbib}
\usepackage{nameref}

\DeclareSymbolFont{symbolsC}{U}{txsyc}{m}{n}
    \DeclareMathSymbol{\lambdaslash}{\mathord}{symbolsC}{110}

\theoremstyle{definition}

\newtheorem*{KB}{Knowledge Balance Principle}
\newtheorem*{defin}{Definition}
\newtheorem*{PC}{Possibilistic Completeness}
\newtheorem*{ROD}{Restcricted Ontic Indifference}

\usepackage{tocloft}

\makeatletter
\renewcommand{\@cftmaketoctitle}{}
\makeatother

\addtocontents{toc}{\cftpagenumbersoff{section}}
\addtocontents{toc}{\cftpagenumbersoff{subsection}}

\newcommand*{\noaddvspace}{\renewcommand*{\addvspace}[1]{}}
\addtocontents{toc}{\protect\noaddvspace}

\begin{document}

\begin{frontmatter}

\title{\textbf{$\boldsymbol{\psi}$-Epistemic Models, Einsteinian Intuitions, and No-Gos\\ \normalsize{A Critical Study of Recent Developments on the Quantum State}}}

\author{Florian Boge}
\address{University of Cologne, Philosophy Department, \\
University of D\"usseldorf, Philosophy Department/DCLPS \\
\href{mailto:boge@phil.hhu.de}{boge@phil.hhu.de}
}




\begin{abstract}
Quantum mechanics notoriously faces the \emph{measurement problem}, the problem that if read thoroughly, it implies the nonexistence of definite outcomes in measurement procedures. A plausible reaction to this and to related problems is to regard a system's quantum state $\ket{\psi}$ merely as an indication of our lack of knowledge about the system, i.e., to interpret it \emph{epistemically}. However, there are radically different ways to spell out such an epistemic view of the quantum state. We here investigate recent developments in the branch that introduces hidden variables $\lambda$ in addition to the quantum state $\ket{\psi}$ and has its roots in Einstein's views. In particular, we confront purported achievements of a concrete model that has been considered to serve as evidence for an epistemic view of the envisioned kind, as well as specific no-go results and their import. It will be argued that while an epistemic account of the particular kind is not straightforwardly ruled out by the no-go results, they demonstrate that the evidential character of the model(s) discussed rests on a rather shaky foundation, and that they make some achievements widely recognized in the literature appear worthy of doubt. 
\end{abstract}


\end{frontmatter}

\section{Introduction}
Quantum mechanics (QM), construed broadly, is the scientific theory with the greatest practical impact and predictive success (cf.~e.g.~\cite[p.~116]{rosenblumkuttner2011} or \cite[p.~893]{kleppnerjackiw2000} for examples), and yet to date it is still faced with the infamous \emph{measurement problem} (MP) -- the problem that the unitary time evolution allows for and preserves superposition states and thus provides no dynamics that lead to definite outcomes in measurement procedures -- and with related issues, all ultimately rooted in quantum superposition. Dirac \cite[p.~7]{dirac1958} and von Neumann \cite[p.~217]{vonneumann1955} historically attempted to solve the MP by adding the \emph{projection postulate} (PP) to the theory, which says that when observable $A$ is measured on system $S$ in state $\ket{\psi(t)}$, the state of $S$ undergoes a sudden change $\ket{\psi}\longrightarrow \frac{\hat{P}_{a}\ket{\psi}}{\|\hat{P}_{a}\ket{\psi}\|}$ with probability $|\braket{a}{\psi(t)}|^{2}$. Here $\hat{P}_{a} = \dyad{a}$ is the projection operator onto the subspace spanned by $\ket{a}$, so upon conclusion of this process, the sate of the system is the (normalized) eigenstate $\ket{a}$ of $A$. In modern QM, this theme readily generalizes\footnote{This is not to say that L\"uders' rule is the general rule to describe state transformations; it rather ``characterizes just one (albeit distinguished) form of state change that may occur in appropriately designed measurements of a given observable with a discrete spectrum.'' \cite[p.~356]{buschlathi2009}} to \emph{L\"uders' rule} $\hat{\rho} \longrightarrow \frac{\hat{M}_{m}\hat{\rho}\hat{M}^{\dagger}_{m}}{\mathrm{Tr}(\hat{M}_{m}\hat{\rho}\hat{M}^{\dagger}_{m})}$, where $\hat{\rho} = \sum_{j} \lambda_{j} \dyad{\psi_{j}}$ is the system's density matrix that may represent a \emph{mixed} quantum state (in case there are two $\lambda_{j}\neq\lambda_{k}$ which are both unequal to zero), and where  $\lbrace\hat{M}^{\dagger}_{m}\hat{M}_{m}\rbrace_{m\in I}$ ($I$ some indexing set) is some more general \emph{positive operator valued measure} (POVM). Many POVMs result from coarse-graining projector valued measures (PVMs) \cite[p.~3]{busch1995}, and this ``may or may not admit the kind of ignorance interpretation familiar from classical physical experimentation.'' [ibid.] The use of mixed states implies that no \emph{pure} quantum state is assigned even beforehand, and (despite the non-uniqueness of decomposition of a mixed state) in particular cases this may be understood as an expression of the fact that the `true' quantum state is simply unknown \cite[cf.~e.g.][p.~20]{mittelstaedt1996}. To that extent generalized measurements may smell like epistemic uncertainties being crucially involved in measurements in QM, and state changes being `merely informational' in some sense. It is not clear though that these considerations carry over to other types of measurement (or even to all generalized measurements), and on account of the \emph{eigenvalue-eigenstate link}, i.e., that an observable $A$ \emph{has} value $a$ on a system $S$ \emph{iff} the state of $S$ is given by an eigenvector $\ket{a}$ of the operator $\hat{A}$ representing $A$, the `sudden change' mentioned above would clearly have to be interpreted differently.

The traditional Dirac-von Neumann approach has evidently raised many questions as to what even \emph{constitutes} a measurement though, and it has (among other things) spawned off fanciful interpretations in which conscious observation has a direct impact on physical reality (cf.\! \cite{wigner2004}; \cite{londonbauer1939}). Of course today we have a whole range of alternative responses to the MP such as the may worlds interpretation, Bohmian mechanics, objective collapse theories... and what have you, all with their particular vices and virtues. There is, however, one particular response that sticks out in its `naturalness', and it is certainly endorsed in some form or other by ``[t]he philosopher in the street, who has not suffered a course in quantum mechanics'' (Bell's phrase \cite{bell1981}) and, we may add, by many a physicist in the lab who has not concerned himself with the foundations of QM. This response is to deprive even pure quantum states of their ontological significance, and to construe the theory \emph{not} as a description of the behavior of physical systems, but rather as a representation of the \emph{knowledge} an actual or ideal observer or agent can have about these. On such a view, the need for an instantaneous reduction of the state vector upon certain kinds of `measurement-like' interactions is removed at once, and the `collapse' appearing in the PP indeed comes out just a sort of informational update for the experimenter upon registration of a given result.

Interpretations of this general sort are typically called \emph{epistemic} or $\psi$-\emph{epistemic}. However, one can spell out an epistemic interpretation in multiple different ways, with strongly diverging underlying assumptions. Leifer for instance maintains that

\begin{quote}it is important to distinguish two kinds of $\psi$-epistemic interpretation. The most popular type are those variously described as anti-realist, instrumentalist, or positivist. [...] The second type of $\psi$-epistemic interpretation are those that are realist, in the sense that they do posit some underlying ontology. They just deny that the wavefunction is part of that ontology. Instead, the wavefunction is to be understood as representing our knowledge of the underlying reality, in the same way that a probability distribution on phase space represents our knowledge of the true phase space point occupied by a classical particle. \cite[p.~72]{leifer2014} \end{quote}

We are here concerned with epistemic interpretations of the second type in Leifer's classification, the kind of interpretation of quantum states which is realist in a decisive sense and arguably strives for as much preservation of \emph{common sense} as possible. The decisive sense of realism here is at least a \emph{metaphysical} or \emph{external} one, meaning that ``[t]he world is (largely) made up of objects that are mind-, language-, and theory-independent.'' \cite[p.~8]{button2013} According to such a realism, there should be no doubt that every system always has a unique, definite state, a unique way of how it `actually is', despite our ignorance of this `actual how'. This should also---a slightly stronger assumption---at least \emph{in principle} enable us to give a unequivocal description of that definite state. The `weirdness' that QM is notoriously associated with is just an expression of our inability to properly \emph{access} the true states of certain (typically microscopic) systems, and hence it vanishes when properly construed in terms of incomplete knowledge.

\section{Einstein's Views and Hidden Variables}\label{sec:ensemble}
A central tenet underlying this type of epistemic interpretation is that QM is in fact an \emph{incomplete} theory that will (hopefully) be replaced by a more complete and comprehensive one in the future. This assumption surfaced early on when the peculiar features of QM became apparent, and its most prominent proponent was, of course, Einstein. This is most vividly reflected in his 1939 correspondence with Schr\"odinger, where he writes:

\begin{quote}I am as convinced as ever that the wave representation of matter is an incomplete representation of the state of affairs, no matter how practically useful it has proved itself to be. The prettiest way to show this is by your example with the cat (radioactive decay with an explosion coupled to it.) At a fixed time parts of the $\psi$-function correspond to the cat being alive and other parts to the cat being pulverized.

If one attempts to interpret the $\psi$-function as a complete description of a state, independent of whether or not it is observed, then this means that at the time in question the cat is neither alive nor pulverized. But one or the other situation would be realized by making an observation.

If one rejects this interpretation then one must assume that the $\psi$-function does not express the real situation but rather that it expresses the contents of our knowledge of the situation. \cite[p.~43]{einstein2011}\end{quote}

Due to Einstein's brave advocacy of these views, we here coin the kind of epistemic interpretation in question \emph{Einstein epistemic} (EE). Due to Bohr's immortal influence on the other sort of epistemic interpretation (the first kind in Leifer's classification), we will call it \emph{Bohr epistemic} (BE), in contrast. We emphasize that an EE account need not encompass \emph{all} of Einstein's views on the quantum state; we here merely use his core intuition, that ``the $\psi$-function does not express the real situation but rather [...] the contents of our knowledge'' to form a label.\footnote{It will be interesting though to discuss at least to some extent which of Einstein's intuitions can or cannot not be preserved.}

Ironically, young \emph{Heisenberg} also spoke merely of a ``destruction of the \emph{knowledge} of a particle's momentum by an apparatus determining its position'' \cite[p.~21; my emphasis -- FB]{heisenberg1930}, and while he himself went on to develop ontologically more charged philosophical views of the quantum state, according to an EE view this is basically \emph{all there is} to most if not all of QM's weirdnesses.

But how does one spell out such an EE view in detail? Einstein's writings are often associated with an \emph{ensemble interpretation} of quantum states because of his continuing appeal to statistical ensembles, as witnessed, e.g., in his reply to criticisms in Schlipp's volume on his life and work \cite[cf.][p.~668]{einstein1949}, or in his 1936 \emph{Physics and Reality}, where he writes: ``The $\psi$-function does not in any way describe a condition which could be that of a single system; it relates rather to many systems, to `an ensemble of systems' in the sense of statistical mechanics.'' \cite[p.~375]{einstein1936}

While there remains some controversy over what Einstein's use of the word `ensemble' actually entailed (cf. \cite{fine1984}; \cite[p.~239~ff.]{whitaker1996}), a more explicit view of this kind was later defended and extended by Ballentine, who describes it in the following terms: 
\begin{quote} For example, the system may be a single electron. Then the ensemble will be the conceptual (infinite) set of all single electrons which have been subjected to some state preparation technique (to be specified for each state), generally by interaction with a suitable apparatus. Thus a momentum eigenstate (plane wave in configuration space) represents the ensemble whose members are single electrons each having the same momentum, but distributed uniformly over all positions. \cite[p.~361]{ballentine1970}\end{quote} 

The ``state preparation technique'' must in fact rather be viewed as an \emph{equivalence class} thereof \cite[cf.][p.~5]{busch1995}, since it is possible to use either a calcite crystal or a grid polarizer, say, to prepare a photon in a certain polarization state, and since different sets of preparation procedures can be used to prepare the same mixed state. 

It may not be immediately obvious why this view of the quantum state should count as `epistemic', but it is obvious from the above quote that at least Einstein was aiming for an epistemic interpretation. He may, in fact, have equally had in mind an \emph{ideal} or \emph{conceptual} ensemble, as suggested in Ballentine's quote, construed as a cognitive tool for determining probabilities of experimental outcomes. This is also the viewpoint of Harrigan and Spekkens \cite[p.~150]{harriganspekkens2010} who think that 
``the ensembles Einstein mentions are simply a manner of grounding talk about the probabilities that characterize an observer's knowledge [...]'' and similarly of Bartlett et al. \cite[p.~4]{bartlett2012} who believe that ``the thesis that quantum states describe the statistical properties of a virtual ensemble of systems [...] is equivalent to saying that it describes one's limited information about a single system drawn from the ensemble.'' 

But \emph{what} exactly \emph{do} these ensembles consist of? They cannot consist of conceptual electrons (say) \emph{in the sense of QM}, because all QM assigns is the state vector which does not---and \emph{cannot}---attribute definite values for all observables at all times. So the conceptual ensembles must consist of something else, something not exhaustively described by QM, but only by some set of additional \emph{hidden variables}. 
 
Epistemic hidden variable approaches of course have a long history in QM as well, and they are the most explicit version of the general contention that QM is incomplete and that there are additional features in nature, more in line with the concepts of classical physics and everyday life thinking (cf.\! also \cite[p.~xvii~ff.]{belinfante1973}). The assumption of hidden variables is also clearly presupposed by ensemble approaches such as Ballentine's, as is evident from Whitaker's analysis of them:

\begin{quote}[I]f the wave-function of a free particle is an eigenfunction of momentum, all members of the ensemble will have the corresponding value of momentum, but in addition each has a precise value of position, though these values will all be different. The values of position must be called hidden variables, because they are not related to the wave-function. \cite[p.~210-211]{whitaker1996}\end{quote}

In fact, all \emph{meaningful} ensemble interpretations of QM have to assume hidden variables (as has also been cogently pointed out by Home and Whitaker \cite[p.~263~ff.]{homewhitaker1992} and d'Espagnat \cite[p.~297~ff.]{despagnat1995}). But there is also the rather trivial sense in which the term `ensemble' plays a role in QM proper: since QM is concerned with probabilistic predictions, the state function must always \emph{also} be allowed to refer to an entire ensemble of equally prepared systems \cite[cf.][p.~6]{belinfante1973}. If QM is considered to be complete however, then $\psi$ should be taken to represent, at the same time, the state of an \emph{individual} system, which is decidedly \emph{not} the case with probability distributions (or densities) used in classical statistical mechanics. 

The traditional ensemble interpretations are \emph{conceptually revisionary} in this sense, in enforcing a break with the assumptions underlying orhtodox QM. But they are \emph{formally conservative}: QM \emph{as it is} may serve as a formal tool for devising statistical predictions; the possibility of a more complete physical theory endorsing hidden variables in a formally \emph{explicit} sense is merely left open or hoped for. 

There are, however, well known reasons why such formally conservative epistemic interpretations never really took off. Schr\"odinger \cite[p.~156]{schrodinger1983},\footnote{Cf.\! also \cite[p.~214]{whitaker1996} for a more detailed analysis of Schr\"odinger's examples.} for instance, started developing counter-examples early on. One of his examples was a harmonic oscillator with a given fixed value of total energy ($(n+\frac{1}{2})\hbar\omega$ for some fixed $n$, say). In the corresponding quantum state of the system, an eigenstate of energy, there would be a large uncertainty as to the oscillators position $\bm{x}$. But according to an ensemble view, kinetic \emph{and} potential energy, depending on velocity and position respectively, should be well defined at all times for each individual member of the ensemble (or rather: the physical objects referred to), whence there should be a clear cut-off value for the position. To see this, consider that the potential energy increases with position $\bm{x}$ in the oscillator potential, and kinetic energy (increasing with velocity) cannot be negative (i.e., the oscillator does not `less than not move'), so that the limit $(n+\frac{1}{2})\hbar\omega$ in total energy implies a limit for possible positions. But QM of course predicts non-vanishing probabilities for positions beyond that limit. 

Similar worries are raised by quantum tunneling. In $\alpha$-decay, an $\alpha$-particle has to tunnel through the Coulomb barrier of the nucleus in order to be emitted. This escape is impossible in a classical physical scenario, since the particle would have to have greater potential than total energy at some point, which again implies negative kinetic energies \cite[cf.][pp.~214-215]{whitaker1996}. 

And finally, arguments that appeal to quantum \emph{interference} are typically invoked, since a simple statistical particle interpretation does not predict the observed interference patterns in double slit experiments and the like. Only the incorporation of an active part of the diaphragm (and possible detectors behind both slits) may raise hopes for a suitable statistical analysis in terms of ensembles. Nothing of this sort is present in the formally conservative ensemble interpretations.  

In summary, what these and many further examples show is ``how far away from the basic [...] ensemble one has to go -- [...] as Bohr would have stressed, one must include the measuring device as an active participator in the measurement, not just a recorder of a fixed value.'' \cite[p.~217]{whitaker1996} And the failure to do so may be seen as the major crux of the historical ensemble approaches. 

\section{Formal Revisions and Epistemic Models}
Modern epistemic approaches in fact \emph{do} use a revised formal inventory, including the possibility for an active part of the measuring device in producing the outcome statistics. From this, they can \emph{prima facie} successfully reproduce many predictions peculiar to QM from merely  epistemic restrictions -- including examples of the infamous quantum interference phenomena.

A particularly influential formal framework that allows to formulate an epistemic view of quantum states in more detail, developed originally by Spekkens \cite{spekkens2005} and extended in joint work with N. Harrigan \cite{harriganspekkens2010}, and very much like (if not an instance of) the formalism used by Bell \cite{bell1964}, is that of the so called \emph{ontological models} (OMs).

\subsection{The Ontological Models-Approach: General Outline}\label{sec:OntMod}
To define what an OM is, Harrigan and Spekkens presuppose an \emph{operationalistic} understanding of QM, which in their words means that ``the primitives of description are simply preparation and measurement procedures -- lists of instructions of what to do in the lab'' \cite[p.~128]{harriganspekkens2010}, and they contend that the goal of this operational formulation is just to determine outcome probabilities for measurement procedures. In contrast, the primitives of description in an OM for this operational theory are the properties of microscopic systems (ibid.). 

To match the operational reading with the quantum formalism, they associate a \emph{preparation procedure} $P$ with a density operator $\hat{\rho}$ (whereas Spekkens \cite[p.~3]{spekkens2005}, more accurately, associates an \emph{equivalence class} of such with $\hat{\rho}$) and a \emph{measurement} $M$ with a POVM $\lbrace\hat{E}_{j}\rbrace_{j\in J}$. But how, precisely, do we have to understand the `association'? What \emph{exactly} does the quantum state represent about the system? The preparation procedure itself (or rather: its equivalence class)? That which results from it? A virtual ensemble in the sense of section~\ref{sec:ensemble}? In contrast to Harrigan and Spekkens, Busch et al., for instance, write: ``Any type of physical system is characterised by means of a collection of preparation procedures, the application of which prepare the system in a \emph{state} $T$. The set of states is taken to be convex, thus accounting for the fact that different preparation procedures can be combined to produce mixtures of states.'' \cite[p.~5, emphasis in original]{busch1995} Here the state is named \emph{in addition to} the collection of preparation procedures, as that which results from them. We will here hence make sense of the association as follows: The quantum state $\hat{\rho}$ of a given system is the state of the system \emph{according to its preparation}, whence we may read the word `state' in a decidedly non-ontological fashion: it does not represent how the system actually \emph{is}, but rather what can be \emph{said} about it, in virtue of what was \emph{done} to it. $\hat{\rho}$ may of course be a pure state and correspond to an eigenstate of some operator, whence the eigenvalue-eigenstate link is clearly severed in this approach. 

\emph{Outcomes} in \emph{projective} measurements, however, can be identified with (pure) quantum states as well, whence quantum states should also be allowed to represent `states according to measurement'. Notably, this reading fits well with the general operationalism about QM, since the measurement is also an operation performed on the system and generally not so much different from the preparation procedure (think of a Stern-Gerlach measurement, where both preparation and measurement involve magnets and screens). Indeed, we thus preserve \emph{half} of the eigenvalue-eigenstate link, i.e.\! that when an observable $A$ is measured to have value $a$ on $S$, the state of $S$ is given by $\ket{a}$---though only in the operational reading of `state'. 

In accord with our analysis, we will, in what follows, occasionally call $\hat{\rho}$ (or $\psi$) the \emph{P/M-state} of a system. Given this understanding of quantum states,  Born's rule provides a probability $\mathrm{Pr}_{M}^{\hat{\rho}}(k) = \mathrm{Tr}(\hat{E}_{k}\hat{\rho})$ of obtaining value $k$ in a \emph{measurement} procedure of type $M$ given some \emph{preparation} procedure resulting in $\hat{\rho}$, i.e., with the meanings of the indices of the probability function loosened. 

To define the notion of an OM, Harrigan and Spekkens now introduce further formal inventory. The first ingredient is a state space $\Lambda$ with elements $\lambda$, termed \emph{ontic states}. These ontic states are supposed to represent a ``complete specification of the properties of a system [...].'' \cite[p.~128]{harriganspekkens2010} Talk of `ontic' states, however, seems somewhat clumsy, whence we will prefer to speak of \emph{true} states instead, i.e.\! states which are true of the systems under consideration, in a correspondence theoretic understanding of truth. This is what the OM approach (obviously) aims for. 

In addition to the space of true states $\lambda$, two probability densities are defined. The first one is termed \emph{epistemic state}, and is intended to reflect the knowledge a possible observer might have about the $\lambda\in\Lambda$. Hence it corresponds to a conditional probability $p(\lambda|\hat{\rho})$ or $p_{\hat{\rho}}(\lambda)$ of obtaining a certain true state $\lambda$, conditional on having prepared P-state $\hat{\rho}$. The second one, denoted by $p(k|\lambda, M)$ or $\xi_{M}^{k}(\lambda)$, is called an \emph{indicator-} or \emph{response function},\footnote{Response functions in statistical mechanics need not be normalized, and therefore need not be interpretable as probability densities in general. In the present context, we assume this to be the case though, in accord with the literature on the subject studied here.} and it is supposed to reflect uncertainties in a given measurement $M$ leading to outcome $k$, conditional on the fact that state $\lambda$ obtains on a system \cite{spekkens2005, harriganspekkens2010}. 

To formally connect the probabilities occurring in the OM approach to the quantum probabilities, Harrigan and Spekkens \cite[p.~128]{harriganspekkens2010} require that an OM must respect the constraint
\begin{equation}\label{eq:OMProb}
\int \dd{\lambda} \xi_{M}^{k}(\lambda)p_{\hat{\rho}}(\lambda) = \mathrm{Tr}(\hat{E}_{k}\hat{\rho}), \ \forall \hat{\rho},M
\end{equation}
to be a model of QM.  That is: adding up (integrating) all the probabilities of obtaining a given outcome $k$, given a certain measurement $M$ and true state $\lambda$, weighted by the probability that the state $\lambda$ even occurs after having prepared $\hat{\rho}$, must reproduce the quantum probabilities.\footnote{For the mathematically inclined, measure theoretic generalizations can be found in \cite[p.~82]{leifer2014}.} This \emph{fully} defines what an ontological model \emph{is},\footnote{\label{fn:TransMat}In \cite{spekkens2005}, the possibility of including a transformation matrix $\Gamma(\lambda, \lambda ')$ in \eqref{eq:OMProb} is discussed, to the end of analyzing the notion of contextuality within the OM approach. We will not dedicate much attention the these here though, since contextuality is not a major subject of the paper.} the state space $\Lambda$, the epistemic state and the response function, and the connection to QM given by formula~\eqref{eq:OMProb}. Hence `ontological model' should be read here rather as a \emph{technical term}; not much ontology is actually conveyed. The OM approach provides a formal framework for analyzing different interpretations of QM, and sketches a road to modifications of QM's formalism which \emph{allow} for the specification of an ontology in which the quantum state does not figure, or at least not fundamentally. It does not yet provide such an ontology.

\begin{figure}
	\centering
	\includegraphics[scale=0.55]{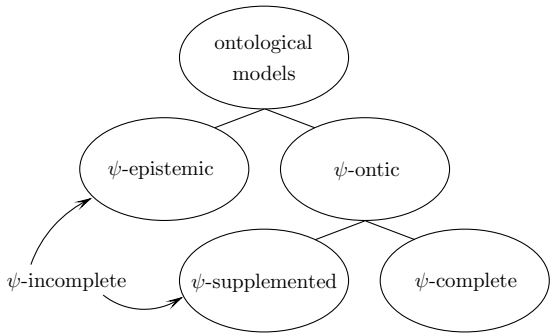}
	\caption[Ontological models]{\small{Classification of OMs according to the status of the quantum state.}}
	\label{fig:OMclass}
	\end{figure}

\emph{Ipso facto}, we are here dealing with an \emph{explicit} hidden variables-approach, the true states $\lambda$ being the hidden variables. Generally $\lambda$ \emph{need not} be interpreted as a hidden variable though, since it can be interpreted as the quantum state $\psi$ itself---the OM-approach is \emph{formally} neutral on this point. In fact, Harrigan and Spekkens \cite[p.~129~ff.]{harriganspekkens2010} draw a multifold distinction between classes of OMs (cf.\! figure~\ref{fig:OMclass}), with a dichotomy of $\psi$-\emph{onitc} and $\psi$-\emph{epistemic} models. Intuitively this means that the quantum state can either be construed as something that actually pertains to real, mind-independent systems, or instead something we ascribe to those systems only in virtue of our lack of knowledge about their true states. Within the first category they again distinguish $\psi$-\emph{supplemented} from $\psi$-\emph{complete} models, where the former category simply consists of OMs in which the quantum state is something that pertains to reality, but still not \emph{all} there is. Intuitively, \emph{Bohmian Mechanics} is an example of such a `model', though it is not clear that it formally fits the approach \cite[cf.][]{feintzeig2014}. The notion of $\psi$-\emph{complete} models should now be self-explaining. For obvious reasons, $\psi$-epistemic and $\psi$-supplemented models are jointly termed $\psi$-\emph{incomplete}.

To get a hold on the more precise definitions of the different subclasses of models, it is sufficient to look into the definitions Harrigan and Spekkens provide for $\psi$-onticity and $\psi$-completeness, as the other concepts can be stated in terms of negations of these.
\begin{defin}[$\psi$-completeness] An ontological model is $\psi$-complete if the space of true states $\Lambda$ is isomorphic to the projective Hilbert space $\mathcal{P(H)}$ (the space of rays of Hilbert space) and if every preparation procedure $P_{\psi}$ associated in quantum theory with a given ray $\psi$ is associated in the OM with a Dirac delta function centered at the true state $\lambda_{\psi}$ that is the value of $\psi$ in the isomorphism, $p_{\psi}(\lambda) = \delta(\lambda-\lambda_{\psi})$.\footnote{Cf.\! \cite[p.~131]{harriganspekkens2010}. We have slightly altered the wording in the definition, as Harrigan and Spekkens call $\lambda_{\psi}$ and $\psi$ isomorphic, but it is meaningless to talk of \emph{elements} of spaces as `isomorphic'. The appeal to projective space Hilbert space here is due to the invariance of quantum states under multiplication by a global phase.}  
\end{defin}
Put frankly, this definition tells us that an OM is $\psi$-complete in case it reproduces QM \emph{tout court}. The true states in $\Lambda$ are  bijectively mapped onto rays in Hilbert space, and the probability of a true state obtaining, given a preparation procedure associated with a ray in $\mathcal{H}$, is such that it is 1 for the true state that is the value of the ray in the isomorphism, and zero for all other true states. The quantum statistics is reproduced in a trivial fashion. 

As we saw, the notion of $\psi$-onticity is supposed to allow for supplementation of $\psi$ by elements of the model which do not simply mirror elements of QM, whence a $\psi$-ontic model is defined as follows.
\begin{defin}[$\psi$-onticity] An ontological model is $\psi$-ontic if for any pair of preparation procedures, $P_{\psi}$ and $P_{\phi}$, associated with distinct quantum states $\psi$ and $\phi$, we have $p_{\psi}(\lambda)p_{\phi}(\lambda) = 0$ for all $\lambda$.\footnote{Cf.\! \cite[ibid]{harriganspekkens2010}.}
\end{defin}

In other words, the supports of two epistemic states that are conditional on two different procedures for preparing distinct P-states should not overlap (on sets of non-zero measure). Now $\psi$-ontic models which do not satisfy the first definition are called $\psi$-supplemented, non-$\psi$-ontic models are called $\psi$-epistemic. The decisive criterion for a model to be $\psi$-epistemic is hence that there be an overlap in (the supports of) the epistemic states associated with distinct quantum states. The intuition being that, if it may happen that $\lambda$ is really the case in two instances, and we have prepared for $\psi$ in the one and for $\phi$ in the other, then $\psi$ and $\phi$ themselves do not reflect something pertaining to the system. Given that the probability distributions $p_{\hat{\rho}}(\lambda)$ were supposed to represent ``what can be known and inferred by observers'' \citep[p.~129]{harriganspekkens2010}, this can be translated more crisply into the statement that one cannot know/infer for sure that a given $\lambda$ is not sometimes the case when one prepares for $\psi$, and sometimes when one prepares for $\phi$. 

Note that sometimes modifications to the definition of $\psi$-epistemicity are discussed \cite[e.g.][]{maroney2013} such that an overlap is only required for \emph{non-orthogonal} states, since two orthogonal states $\ket{\phi}, \ket{\psi}$ may be construed as indicative of mutually exclusive preparation procedures that could arguably result in mutually exclusive sets of true states, or even that \emph{all} non-orthogonal states should be associated with overlaps \cite[e.g.][]{lewisjennings2012}. We here however stick to the definition given above, and the negation of $\psi$-onticity here merely implies the \emph{existence} of two such distinct (possibly non-orthogonal) states that have distributions associated to them with overlapping supports. This is a comparatively weak requirement, and refinements in terms of distance measures between the epistemic states have also been proposed (see e.g.\! \cite[p.~477]{pbr2012}; \cite[p.~2]{aaronson2013}). 

Again, we find justification for associating this kind of epistemic interpretation with Einstein, since the idea of demonstrating the incompleteness of QM in virtue of the existence of two $\psi$ functions that correspond to the same real physical state of an object at the same time was also explicitly advocated by him:\footnote{This is also a key reason why Harrigan and Spekkens \cite{harriganspekkens2010} devote a large part of their paper to Einstein's views of the quantum state.} ``[...] co\"ordination of several $\psi$ functions with the same physical condition of [some] system [...] shows [...] that the function cannot be interpreted as a (complete) description of a physical condition of a unit system.'' \cite[p.~376]{einstein1936} We repeat, however, that the OM approach \emph{in general} merely constitutes a formal framework for analyzing different interpretations of QM or of the quantum state. Here we are only interested in the approach to the extent that it can accommodate the intuitions underlying an EE view, i.e.: \emph{we are here only concerned with $\psi$-epistemic models, and we take them to be a suitable implementation of an EE view.}

\subsection{Interlude: The Philosophical Issues at Stake}\label{sec:PhilPresOM}
What is at stake in the present discussion of the quantum state? In fact, there is a whole range of topics that could be picked out, but we will focus on three concerns and how they relate to the debate.

The search for `completeness' is our first concern, and it raises some first worries as regard the definitions at hand. Recall that $\lambda$ was supposed to correspond to a ``complete specification of the properties of a system[...].'' But on a broad reading of `property' this seems rather impossible. For suppose that we have prepared for $\psi$ in the one case and $\phi$ in the other, and that in both cases $\lambda$ is supposed to occur. Then we can say that there is a (complex) \emph{relational} property of being-in-a-$P_{\psi}$-situation in the first case, and a relational property of being-in-a-$P_{\phi}$-situation in the second, and hence, $\lambda$ cannot strictly specify \emph{all} properties.

Bell, whose work has certainly served as a paradigm for the OM approach (as noted above), more cautiously talked about a ``\emph{more} complete specification'', for which it is ``a matter of indifference [...] whether $\lambda$ denotes a single variable or a set, or even a set of functions, and whether the variables are discrete or continuous.'' \cite[p.~15; my emphasis -- FB]{bell1964} Clearly Harrigan and Spekkens must have in mind a particular set of \emph{physical} or \emph{natural} properties; kinematical quantities comparable to those arising from the phase space formalism in classical mechanics, i.e.\! all (and only) those properties that can be defined in terms of (generalized) positions and momenta. Ruetsche \cite[p.~31]{ruetsche2011} in fact makes a point in our favor, in identifying generalized coordinates and momenta as paradigmatic examples of magnitudes she calls ``fundamental in the physicist's sense,'' meaning that it is usually assumed that the value of \emph{every} other \emph{magnitude} pertaining to a system can be determined by assigning values to these on the system. The `completeness' sought for by Harrigan and Spekkens must consist in seeking such fundamental-in-the-physicist's-sense quantities. Quantum mechanically the closest analogy is the complete set of comeasurable observables. But since these of necessity only include a subset of all conceivable observables, the contention (again) is that there must be other, hitherto undiscovered magnitudes \emph{beyond} QM's observables that are more in line with the classical description. 

But one should ask what actually singles out such a set of properties, beyond their \emph{usefulness} for the applicability of a given theory. How can we get hold of a preferred set of `natural' physical properties which `carve nature at its joints', as it were, and suffice to completely describe a system in an ontologically privileged way? As regards usefulness, QM proper arguably has quite an edge in the light of all its predictive and technological successes---\emph{despite} its lack of definite descriptions similar to the those occurring in classical theories. So seeking completeness in more than a theory-internal sense first of all requires a sensible criterion for naturalness of properties. 

Besides the completeness-issues, a second point to wonder is what concept(s) of \emph{probability} are involved in $\psi$-epistemic models. This is a point which Harrigan and Spekkens refrain from elucidating, as they ``do not feel that the distinction [between different concepts of probability -- FB] is significant in this context [...].'' \cite[p.~150]{harriganspekkens2010} This distinction may not be as insignificant as they apparently believe though, since (famously) there is a whole host of radically differing views of probability, and opting for one particular view always comes with rather deep ontological and epistemological implications. Possibly the broadest dichotomy one can draw is that between \emph{epistemic} and \emph{objective} probabilities \cite[cf.~e.g.][p.~2]{gillies2000}, and on this coarse level the classification of the epistemic state seems pretty clear. But as regards the response function, matters are more subtle. Harrigan and Spekkens here have it that ``the model may be such that the ontic state $\lambda$ \emph{determines only} the probability $p(k|\lambda, M)$ of different outcomes $k$ for the measurement $M$.'' \cite[p.~128; my emphasis -- FB]{harriganspekkens2010} And this smells like the response function involving objective probability. But in fact they also claim that ``$p(\lambda|P)$ \emph{and} $p(k|\lambda, M)$ specify what can be known and inferred by observers'' \cite[p.~129; my emphasis -- FB]{harriganspekkens2010}, which \emph{prima facie} conflicts the previous quote. But remember that their claim merely was that an (epistemic) OM leaves it \emph{open} whether (``may be such that'') $\lambda$ determines only outcome probabilities for values $k$ in measurement $M$. And if there is indeed a random response of the measurement device to the true states fed in, then it is, of course, \emph{also} the case that \emph{only} a probability for a given outcome can be known or inferred by an observer. 

For the purposes of an epistemic model, one might still be inclined to hope for an interpretation of the response function according to which randomized responses ``could occur because of our failure to take into account the precise ontological configurations of either [preparation or measurement]'' \cite[p.~4]{harriganrudolph2007}, or simply reflect an ``un\emph{known} disturbance'' \cite[p.~10; my emphasis -- FB]{spekkens2007} of the system caused by the measurement---much like Heisenberg's elimination of knowledge due to the measurement process. However, taking both probabilities to be epistemic in this sense could, firstly, be fleshed out to yield a kind of \emph{microdeterminism}, meaning that on a sufficiently fine-grained scale of observation (accessible only to Laplacian demons, as it were), no probabilities would be needed. This would take the debate to a whole other level though, since currently the central question is one of micro\emph{definiteness} (the \emph{existence} of additional true states which explain the measurement statistics), not determinism. And secondly, we saw that a failure to take the role of the measurement device into account was the main reason for the failure of the historical (ensemble) approaches, so regarding the response function as merely reflecting epistemic uncertainties may not be good advice after all, or at least raises further concerns.

Leifer \cite[p.~70]{leifer2014}, moreover, notes that ``calling a probability density `epistemic' [...] presupposes a broadly Bayes-ian interpretation of probability theory in which probabilities represent an agent's knowledge, information, or beliefs.'' But `broadly Bayesian' may still be too broad in this context, since there is also the \emph{quantum Bayesian} approach, which explicitly endorses \emph{subjective} Bayesianism and refrains from entertaining (formally explicit) hidden variables. Consider Williamson's \cite[p.~iii]{williamson2010} characterization of the distinction between subjective and objective Bayesianism: ``Subjective Bayesians hold that it is largely (though not entirely) up to the agent as to which degrees of belief to adopt. Objective Bayesians, on the other hand, maintain that appropriate degrees of belief are largely (though not entirely) determined by the agent's evidence.'' Moreover, Williamson characterizes objective Bayesianism as a \emph{normative} theory, i.e., a theory which claims that ``[t]he strengths of an agent's beliefs \emph{should} behave like probabilities[...].'' \cite[p.~1; my emphasis -- FB]{williamson2010}
 
\emph{Prima facie} $\psi$-epistemic models, in which the epistemic state quantifies what can be known and inferred by observes, are best construed as embracing an objective Bayesian reading of probabilities. But upon closer inspection, they also appear compatible equally with what Williamson \cite[p.~15]{williamson2010} refers to as \emph{empirically based subjective Bayesianism}. All forms of Bayesianism, Williamson (ibid.) tells us, hold a \emph{Probability norm}, meaning that ``one's degrees of belief at a particular time must be probabilities if they are to be considered rational.'' Empirically based subjective Bayesians add a \emph{Calibration norm}, i.e.\! that ``one's degrees of belief [...] should also be calibrated with known frequencies.'' (ibid.) Because the epistemic state is supposed to reflect what can be known and inferred by an observer (agent) -- on the basis of \emph{evidence} about preparation methods, and the presumed range of true states which can result from each preparation -- we can see that the attribution of probabilities must be on the basis of known frequencies, and the Calibration norm should clearly hold for $\psi$-epistemic models. \emph{Objective} Bayesians additionally assume an \emph{Equivocation norm}, meaning that ``one's degrees of belief at a particular time are rational if and only if they are probabilities, calibrated with physical probability and otherwise \emph{equivocate} between the basic possibilities.'' \cite[p.~16; my emphasis -- FB]{williamson2010} It is not clear that an epistemic OM must embrace this norm as well, but we will investigate a particular model below which clearly does. 

Thirdly, we (re-)acknowledge that these probability densities are probabilities of (or conditional on) \emph{true states} $\lambda$ obtaining, which we above identified as the assumption that makes the whole approach decisively (metaphysically) realist. In fact, all the issues at hand, the status of the epistemic probabilities, the search for completeness, and the status of the quantum state, all boil down to questions of the precise kind of realism endorsed. 

It has been criticized though, in particular by Norsen \citep{norsen2007}, that in certain applications in the context of QM (the violations of Bell-type inequalities) the discussion is blurred by the use of the word `realism', because ``it is almost never clear what exactly a given user means by the term [...] and [...] none of [the] possibly-meant senses of `realism' turn out to have the kind of relevance that the users seem to think they have.'' \citep[pp.~311-312]{norsen2007} Whether Norsen's assessment is correct is open to debate. But we concede that there is a crucial terminological problem that we should try to fix. 

First of all we note that the $\psi$-epistemic OMs need not be considered \emph{na\"ively} realist in the sense appealed to by Norsen \cite[p.~316]{norsen2007}, namely in the sense that ``whenever an experimental physicist performs a `measurement' of some property of some physical system [...] the outcome of that measurement is simply a passive revealing of some pre-existing intrinsic property of the object.'' (emphasis omitted) This,  Norsen thinks, is the physics-appropriate generalization of ``the view that all features of a perceptual experience have their origin in some identical corresponding feature of the perceived object.'' (p.~315) 

\emph{Classical} physics is typically taken to entertain exactly that sort of na\"ive realism, as it seems to endorse that measurements can at least in principle be as subtle and non-invasive as desired (think again of Heisenberg's ``destruction of knowledge'', which he apparently considered as a philosophical revelation). Similarly, the ensemble approaches discussed in section~\ref{sec:ensemble} may be classified as na\"ively realist, but, again, $\psi$-epistemic OMs need \emph{not} be viewed as na\"ively realist in Norsen's sense, since both preparation and measurement are infected with uncertainties, so that pre-existing intrinsic properties of systems are not just revealed passively\footnote{By judging thusly, we are in fact disagreeing with Norsen, who thinks that this na\"ive realism is the idea of a non-contextual hidden variable model. But the disagreement may be based on the understanding of `context'.} -- \emph{at least as long as the response function is interpreted objectively}.

More illuminatingly, it appears that the project of finding $\psi$-epistemic models for QM must embrace a form of \emph{scientific} realism (i.e., that (i) mature and well confirmed theories are \emph{capable} of being true, and (ii) the concepts of these very theories typically \emph{do} refer to entities in the external world, in \emph{all} domains, including unobservable microstates \cite[cf.][p.~xvii]{psillos1999}), since scientific methods are employed to seek out (refer to) the true states of investigated systems. But since QM is interpreted \emph{operationally}, this realism can \emph{only} (or at least: \emph{mostly}) be endorsed here w.r.t.\! \emph{classical} physical concepts and theories.\footnote{Note that, if scientific realism was applied here to the concepts of a hitherto undiscovered physical theory $T$, this would not be more than a mere \emph{hope for} scientific realism. Moreover, $T$ would still have to imply classical physical theories in suitable limits, and since $T$ is decidedly \emph{not} QM, we can take it that an EE view seeks out a $T$ whose concepts would be more clearly compatible with those of classical theories.} The fact that in $\psi$-epistemic models, the aim is to reduce quantum probabilities to classically interpretable ones, and quantum states to definite states that provide a (more) complete specification of reality should suffice as evidence for this claim. Grossly speaking, we can hence classify EE views as \emph{selectively scientific realist}. 

Peters \cite[p.~377]{peters2014} describes selective scientific realism as the view ``that not all the propositions of an empirically successful theory should be regarded as (approximately) true but only those elements that are essential for its success'', but he also notes that ``[i]t is [...] not obvious how a term like `essential' is to be understood.'' If we apply this reading of `selective' to the present case, what would be such essential elements? We may take it that the possibility of a rather gapless (formal) \emph{picture} of reality is considered somewhat essential to scientific theories by advocates of an EE view, since the supplementation of additional variables to the formal inventory (instead of an analysis of the formal elements already present in QM) would otherwise be a moot point. And indeed, Spekkens generally characterizes OMs as ``an attempt to offer an explanation of the success of an operational theory by assuming that there exist physical systems that are the subject of the experiment.'' \cite[p.~2]{spekkens2005} 
 
We can see that rather deep philosophical issues are at stake here, and the OM approach provides not only a technically useful basis for the general discussion of interpretations of QM, but also for the assessment of these very issues in the context of QM. But to assess whether epistemic models formulated in the approach can ascertain the completability of QM, the suitability of objective (or empirically based subjective) Bayesianism for the interpretation of probabilities in QM, and a selectively scientific realist attitude for the general context, we must ask: (i) \emph{Are} there any models which fit the definitions from section~\ref{sec:OntMod}, and (ii) to what extent can these reproduce the empirical predictions of QM? Indeed, Harrigan and Spekkens provide examples of models for each of their categories, but we here turn to a model that Spekkens has devised in 2007 instead, and which, we here take it, has so far brought about the greatest apparent successes in providing evidence for an EE view. We will here grossly focus on this model to evaluate (ii). 

\subsection{Spekkens' Toy Model}\label{sec:toymod}
What we here call a `toy model' was originally developed by Spekkens \cite{spekkens2007} under the name ``toy theory'', but it can be fit into the OM approach, as shown in \cite[p.~84]{leifer2014} and below. The toy model is only concerned with analoga of qubit systems (i.e.\! systems with only two relevant states in QM), but can be expanded to include systems of multiple, coupled qubits. The analogues of qubits in the toy model are called \emph{elementary systems} (cf. \cite[p.~3]{spekkens2007}). For these elementary systems, Spekkens postulates four possible true states, simply denoted by $\qty{1, 2, 3, 4}$. There is a foundational principle at the heart of this model, called the \emph{kowledge balance principle}: 
\begin{KB}[KB]\label{axm:KB}
If one has maximal knowledge, then for every system, at every time, the amount of knowledge one possesses about the [true -- FB] state of the system at that time must equal the amount of knowledge one lacks. \cite[p.~3]{spekkens2007}
\end{KB}

This, of course, immediately raises the question of how to \emph{measure} knowledge. To provide a measure, Spekkens first defines what he calls \emph{canonical sets} (cf.\! ibid.): 
\begin{defin}[Canonical set]\label{dfn:CanoSet}
A \emph{canonical set} is a set of yes-no questions that is sufficient to fully specify the true state, and that has a minimal number of elements. 
\end{defin}

To understand this notion, consider that if one only knows that the state of the system under investigation is in the set $\qty{1,2,3,4}$ and one wants to find out in which of the states it actually is, one could ask ``Is it in state 1?'', ``Is it in state 2?'', and so forth. Or one could be smart instead, and just ask, say, ``Is the system's state in the set $\qty{1,2}$?'', and ``Is the system's state in the set $\qty{2,3}$?'' Two nos will give assurance that it is in state 4, two yeses that it is in 2, and one yes and one no that it is either in 1 or 3, depending on the order. Now the amount of knowledge one has is defined within the toy model as ``the maximum number of questions for which the answer is known, in a variation over all canonical sets of questions.'' (ibid.).

(\nameref{axm:KB}) then dictates that one can always only know half the answers in such a set, and this is somewhat reminiscent of an epistemic reading of the uncertainty relations. Applied to physical systems such as spinful particles, we can understand it such that, ``if we know the $x$-coordinate [of spin -- FB] with certainty then we cannot know anything about the $y$-coordinate.'' \cite[p.~73]{leifer2014} Put frankly, this means that the epistemic states (for simple systems) within this model must be distributions which assign probability $1/2$ to two states, and probability 0 to two others. That is, we can always know that the state is in a subset like $\qty{1,2}$, but nothing more. 

To connect these principles to QM, consider (as does Spekkens) the following six quantum states, which are the only P/M-states of the model:
\begin{align*}
&\ket{0}, & &\ket{1}, \\
&\ket{+} = \frac{1}{\sqrt{2}}(\ket{0} + \ket{1}), & &\ket{-} = \frac{1}{\sqrt{2}}(\ket{0} - \ket{1}), \\
&\ket{+i} = \frac{1}{\sqrt{2}}(\ket{0} + i\ket{1}), & &\ket{-i} = \frac{1}{\sqrt{2}}(\ket{0} -i \ket{1}),
\end{align*}
with $\braket{0}{1} =0$. Accordingly our epistemic states will be of the form $p(\lambda|x), \ \lambda\in\qty{1,\ldots 4}, \ x\in\qty{0,1,+,-,+i,-i}$.  

Since we are only concerned with a discrete set of possible true states, the probability distributions can be represented by $n$-tuples. This also means that condition~(\ref{eq:OMProb}) which connects the QM probabilities with the probabilities in the OM must be changed to a sum:
\begin{equation}
\mathrm{Tr}(\hat{E}_{k}\hat{\rho})=\sum_{\lambda\in\Lambda}p_{\hat{\rho}}(\lambda)\xi^{k}_{M}(\lambda).
\end{equation}
Spekkens also introduces a convenient notation for the epistemic states, which we will equally make use of in what follows. We hence make the following identifications:
\begin{align*}
&p_{0}=(\textstyle{\frac{1}{2},\frac{1}{2},0,0})\leftrightsquigarrow 1\vee 2, & &p_{1}=(\textstyle{0,0,\frac{1}{2},\frac{1}{2}})\leftrightsquigarrow 3\vee 4, \\
&p_{+}=(\textstyle{\frac{1}{2},0,\frac{1}{2},0})\leftrightsquigarrow 1\vee 3, & &p_{-}=(\textstyle{0,\frac{1}{2},0,\frac{1}{2}})\leftrightsquigarrow 2\vee 4, \\
&p_{+i}=(\textstyle{0,\frac{1}{2},\frac{1}{2},0}) \leftrightsquigarrow 2\vee 3, & &p_{-i}=(\textstyle{\frac{1}{2},0,0,\frac{1}{2}})\leftrightsquigarrow 1\vee 4,
\end{align*}
Curvy arrows are used to denote correspondence between different notations, and the disjunctions should be read `merely symbolic' at this point (we will spend a few thoughts on connections to logic below). We have, e.g., $p_{0}(1)=p_{0}(2)=\frac{1}{2}$ and $p_{0}(3)=p_{0}(4)=0$.   

The response functions turn out deterministic here; for instance, 
\begin{equation*}
\mathrm{Pr}_{+/-}^{\ket{0}}(+) = |\braket{+}{0}|^{2} = 1/2 \stackrel{!}{=} \linebreak \sum\limits_{\lambda\in\Lambda}p_{0}(\lambda)\xi_{+/-}^{+}(\lambda),
\end{equation*}
 where `$+/-$' refers to the measurement associated with outcomes $+$ and $-$. But this means that $\xi_{+/-}^{+}(\lambda)$ has to give 1 for the first of the $\lambda$s, and cannot also give 1 for the second one. Equally, 
\begin{align*}
\sum\limits_{\lambda\in\Lambda}p_{1}(\lambda)\xi_{+/-}^{+}(\lambda)\stackrel{!}{=}|\braket{+}{1}|^{2} = 1/2, \\
\sum\limits_{\lambda\in\Lambda}p_{+}(\lambda)\xi_{+/-}^{+}(\lambda)\stackrel{!}{=}|\braket{+}{+}|^{2} = 1, \\ 
\sum\limits_{\lambda\in\Lambda}p_{-}(\lambda)\xi_{+/-}^{+}(\lambda)\stackrel{!}{=}|\braket{+}{-}|^{2} = 0,
\end{align*}
 and so forth. All in all, we get $\xi_{+/-}^{+}(\lambda) = (1,0,1,0)$, so that the $\xi$ for outcome $+$ mirrors the $p$ which is conditional on $+$, but with 1s instead of $\frac{1}{2}$s. All the $\xi$s can be worked out to look this way.\footnote{Note that it is not a contradiction that the entries in $\xi$ sum up to 2 instead of 1, as $\xi$ expressed in this way is variable in $\lambda$, i.e., in the true state on which it is conditional, not in the outcome. Only the sum over all \emph{outcome probabilities}, given \emph{fixed} parameters ($\lambda, M$) must sum to one.}

So $\xi$ actually does not do any work here at all and could be omitted altogether.  Models of this kind have been coined \emph{maximally $\psi$-epistemic} (cf.\! \cite[p.~2]{maroney2013}; \cite[p.~4]{maorenyleifer2013}), and the toy model is such a maximally epistemic model.

With this simple setup, Spekkens is \emph{prima facie} able to reproduce a bunch of quantum phenomena. To this end he assumes measurements to be ``\emph{reproducible} in the sense that if repeated upon the same system, they yield the same outcome.'' \cite[p.~9; emphasis in original]{spekkens2007} In other words: they are like the projective measurements of QM. But as noted before, due to (\nameref{axm:KB}) measurements cannot reveal the true state $\lambda$, but can only change what one knows about the system. To elaborate, first note that a state of total ignorance, where one only knows $\lambda\in\qty{1, 2, 3, 4}$, should be described by an epistemic state $p(\lambda) = 1/4, \forall \lambda\in\Lambda$, or $1\vee 2\vee 3\vee 4$ \citep[p.~4]{spekkens2007}.\footnote{Since nothing at all is known in states like $1\vee 2\vee 3\vee 4$, this can also be construed to mirror completely mixed states that can be decomposed into multiple convex combinations \citep[cf.][p.~5]{spekkens2007}. We here also see that the model \emph{equivocates} between basic possibilities, which substantiates our previous claim that it is objectively Bayesian.}

Upon measurement $1\vee 2\vee 3\vee 4$ will now be changed into a state where one knows one of the (symbolic) disjunctions $1\vee 2, 3\vee 4, 1\vee 3 \ldots$ This is represented in the model as the measurement `inducing a partition', say $\qty{1, 2, 3, 4}\xlongrightarrow{M}\qty{\qty{1, 2}, \qty{3, 4}}$. This amounts to a \emph{probability update}, reminiscent, of course, of Bayesian condtionalization \citep[cf.][p.~75~ff.]{williamson2010}. Let us say that some experimenter has no prior knowledge about the true state of a system, and hence no preference in belief as to which state an investigated system is in. Then her epistemic state should be $p = (\frac{1}{4}, \frac{1}{4}, \frac{1}{4}, \frac{1}{4})$. Upon measuring the value + (say), she will instantaneously think that the system must be in one of the states 1 and 3, but she can still give no preference to any of the two. Thus her knowledge about the system has to be represented as $p_{+} = (\frac{1}{2}, 0, \frac{1}{2}, 0)$. This is the picture provided by the formal setup of the model of what happens in a measurement, and it seems to explain the presence of the projection postulate in orthodox QM.

It is important to note that (\nameref{axm:KB}) is restricted to the knowledge about a system \emph{at a given time}. This is so because given that one knows $1\vee 2$, a measurement which partitions $\qty{\qty{1, 3}, \qty{2, 4}}$ will lead to definite knowledge of the state of the system \emph{prior} to the measurement; in case one measures $1\vee 3$ the state must have been 1, in case of $2\vee 4$ it must have been 2. The fact that one still lacks complete knowledge about the system's state \emph{after} the measurement is accounted for by an ``unknown disturbance'' of the state, caused by the measurement \citep[p.~10]{spekkens2007}. 

The first achievement of this model now is that these measurements can be demonstrated to exhibit \emph{non-}\linebreak\emph{commutativity}, just as quantum measurements do. Consider two measurements $A$ and $B$ inducing partitions \linebreak $\qty{\qty{1, 2}, \qty{3, 4}}$ and $\qty{\qty{1, 3}, \qty{2, 4}}$ respectively, and performed on a system in state $1\vee 2$. Performing the $A$-measurement first will keep the system in $1\vee 2$ and the $B$-measurement will then yield $1\vee 3$ and $2\vee 4$ with equal frequencies. Performing them the other way around, the $B$-measurement will first update the epistemic state to either $1\vee 3$ or $2\vee 4$; but now the $A$-measurement will yield $1\vee2$ and $3\vee 4$ with equal frequency. This should be compared, say, to the non-commutativity of spin measurements along orthogonal axes in a Stern-Gerlach experiment.

\begin{center}
	\begin{figure}
	\centering
	\includegraphics[scale=0.44]{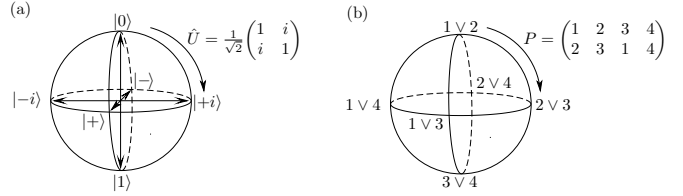}
	\caption[Quantum and epistemc states]{\small{(a) is a regular Bloch sphere for the qubit states, (b) is an analogous diagram for the epistemic states (cf.\! \cite{spekkens2007} for a similar illustration). Two of these can be combined by operations $+_{1},\ldots , +_{4}$ to yield one of the respective other states, just as two quantum states can be superposed to yield a third one. In (a), transformations are represented by unitary operators (which can be mapped onto rotations), in (b) permutations are used instead.}}\label{fig:epibloch}
	\end{figure}
\end{center}

Another achievement is the (partial) reproduction of quantum superposition. This is accomplished by defining different rules for combining the epistemic states. For instance, one could combine two states such as $1\vee 2$ and $3\vee 4$ by taking the true state of lowest index and combining them into a new state, i.e. $1\vee 3$. This may be symbolized by writing $(1\vee 2)+_{1} (3\vee 4)=1\vee 3$. Equally, we could take the true states of highest index to obtain $2\vee 4$, which may be written as $(1\vee 2)+_{2} (3\vee 4)=2\vee 4$. Taking one of higher and one of lower index from both epistemic states respectively will yield two further possibilities ($+_{3}$ and $+_{4}$; cf.\! \cite[p.~6]{spekkens2007}). With these four combination rules, the interrelations of all six quantum sates considered above can be mirrored, which is best illustrated in terms of Bloch spheres (or Bloch sphere-like diagrams), as in figure~\ref{fig:epibloch}.

There are, however, a few subtleties involved in this analogy which lead into a (first) kind of trouble. Combining, say, $(2\vee 3)+_{4}(1\vee 4) = 2\vee 4$ in the toy model should, according to the Bloch sphere-image, be analogous to superposing $\ket{+i}$ and $\ket{-i}$ to get $\ket{-}$ in QM, i.e.\! developing $\ket{-} = \braket{+i}{-}\ket{+i} + \braket{-i}{-}\ket{-i} = \frac{1+i}{2}\ket{+i} + \frac{1-i}{2}\ket{-i}$. A complication is now raised, however, by the fact that combination rules $+_{3}$ and $+_{4}$ have a particular \emph{ordering sensitivity}, as e.g.\! $(1\vee 4)+_{4}(2\vee 3) = 1\vee 3\neq (2\vee 3)+_{4}(1\vee 4)$. One can model this situation by a superposition $\frac{1}{\sqrt{2}}(\ket{+i}-i\ket{-i})$ with relative phase, which is equal to $e^{-i\frac{\pi}{4}}\ket{-}$, because 
\begin{align*}
e^{-i\frac{\pi}{4}} & = \cos(-\frac{\pi}{4}) + i\sin(-\frac{\pi}{4}) = \\
& =  \cos(\frac{\pi}{4}) - i\sin(\frac{\pi}{4}) = \frac{1}{\sqrt{2}}(1-i),
\end{align*}
so that
\begin{equation*}
e^{-i\frac{\pi}{4}}\ket{-}  =  \frac{1}{\sqrt{2}}(1-i)\ket{-} = \frac{1}{\sqrt{2}}(\ket{+i}-i\ket{-i}). 
\end{equation*}
I.e., we obtain $\ket{-}$ up to a(n empirically meaningless global overall) phase, but the superposition rule thus essentially includes a \emph{relative} phase of $\frac{3\pi}{2}$ (since $e^{i\frac{3\pi}{2}}=-i$) between the two states superposed. In fact, the four combination rules above can all be understood in terms of quantum superpositions with a relative phase, and Spekkens \cite[p.~7]{spekkens2007} makes the following identifications: 
\begin{align*}
&+_{1}\leftrightsquigarrow +e^{i\cdot 0}, & &+_{2}\leftrightsquigarrow + e^{i\pi}, \\
&+_{3} \leftrightsquigarrow + e^{i\frac{\pi}{2}}, & &+_{4}\leftrightsquigarrow + e^{i\frac{3\pi}{2}}. 
\end{align*}
These identifications reveal the subtleties mentioned above and show that the analogy between combinations of epistemic states and quantum superpositions  is not---and \emph{cannot} be made---perfect. In the given choice one obtains $(1\vee 3)+_{3}(2\vee 4) = 2\vee 3$ and $(1\vee 3)+_{4}(2\vee 4) = 1\vee 4$, but $\frac{1}{\sqrt{2}}(\ket{+}+e^{i\frac{\pi}{2}}\ket{-})=e^{i\frac{\pi}{4}}\ket{-i}$ and $\frac{1}{\sqrt{2}}(\ket{+}+e^{i\frac{3\pi}{2}}\ket{-})=e^{-i\frac{\pi}{4}}\ket{+i}$, which, given the identifications between combination rules and epistemic- and quantum states, should be exactly the other way around. Exchanging identifications in the latter case will always only shift the problem \cite[cf.][p.~7]{spekkens2007}. According to Spekkens (ibid.), ``[t]his curious failure of the analogy shows that an elementary system in the toy theory is not simply a constrained version of a qubit.'' 

So the toy model fails to correctly reproduce the QM toolkit from epistemic restrictions in this instance, and is bound to do so. But this need not be a strong objection to the general enterprise yet, because (a) we are here dealing with a limited toy model only, and (b) it should not be required that any successful alternative to QM must mirror the quantum \emph{formalism} isomorphically; a successful replacement of, or alternative to QM should only be required to preserve QM's successful \emph{predictions}. If we construe the model, however, as a first approach to \emph{reducing} the exact rules of QM to incomplete knowledge (thereby serving as evidence for an EE view), then it must still appear as a drawback that the model fails to do so.

Be that as it may, further interesting phenomena can (apparently) be reproduced within the toy model in virtue of \emph{state transformations} being represented in the model as \emph{permutations} of true states in the epistemic state, or equivalently, as resamplings of the epistemic state (cf.\! figure~\ref{fig:epibloch}). In the Bayesian paradigm, this amounts to a change in an observer's \emph{knowledge}, the point being that the true state of a system does not have to change, even if the epistemic state of an observer does. One can of course find out some new piece of information, regardless of whether the system this information is about remains entirely unchanged. Intuitively we can access the formal analogy to unitary transformations in the Bloch representation, as the latter correspond to rotations (up to an overall phase) of pointers inside the sphere, and permutations will also appear as rotations by angles of $n\frac{\pi}{2} (n\in\mathbb{Z})$ in the toy sphere (cf.\! figure~\ref{fig:epibloch} (b)). 

\begin{figure}
	\centering
	\includegraphics[scale=0.5]{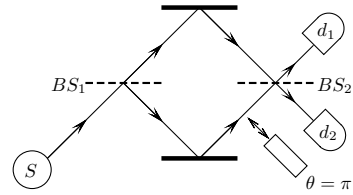}
	\caption[Mach-Zehnder interferometer]{\small{Mach-Zehnder interferometer with an optional phase.}}
	\label{fig:machzehnder1}
	\end{figure}

But what is striking about permutations as state transformations is that they seem to make the reproduction of \emph{quantum interference} examples possible. To elaborate, consider the following setup\footnote{Note that no analogous discussion of such an example is provided in \cite{spekkens2007}.} based on a \emph{Mach-Zehnder interferometer} (figure~\ref{fig:machzehnder1}), where one photon at a time enters the setup, emitted from a source ($S$) towards a 50/50 beam splitter ($BS_{1}$), so that there will be a 50/50 chance for each photon of passing through $BS_{1}$ or being reflected at a right angle. 

Neglecting polarization etc., we can model this as a simple spatial qubit of a `moving up state' $\ket{\nearrow}\doteq{\tiny\mqty(1 \\ 0)}$ and a `moving down state' $\ket{\searrow}\doteq{\tiny\mqty(0 \\ 1)}$ (`$\doteq$' implies a choice of representation). Now take a photon prepared as $\ket{\nearrow}$ by $S$. $BS_{1}$ will change the state into a superposition of moving up and moving down, represented by 
\begin{equation}\label{eq:hadamard}
\hat{U}_{H}\ket{\nearrow} = \frac{1}{\sqrt{2}}(\ket{\nearrow} + \ket{\searrow}) =: \ket{\psi},
\end{equation}
$\hat{U}_{H}\doteq\textstyle{\frac{1}{\sqrt{2}}{\tiny\mqty( 1 & 1 \\ 1 & -1)}}$ the \emph{Hadamard gate}. 

Imagine now that behind each beam of photons emanating from $BS_{1}$ there are mirrors (the thick black lines in figure~\ref{fig:machzehnder1}), aligned such that both trajectories are deflected towards each other again. We can represent the transformation effected by the mirrors by the $\hat{\sigma}_{x}$ Pauli-matrix, which will only exchange the flying up- and down-components of $\ket{\psi}$ and hence essentially leave it untouched, as 
\begin{equation*}
\mqty(0 & 1 \\ 1 & 0)\frac{1}{\sqrt{2}}\mqty(1 \\ 1) = \frac{1}{\sqrt{2}}\mqty(1 \\ 1).
\end{equation*}
We can also insert a phase shifter in the lower branch, say, but after the mirrors, so that it will only affect the flying-up part of the spatial superposition state. If we choose $\theta = \pi$ as our phase, we will obtain a transformation which can be represented by the matrix 
\begin{equation}
\hat{\Phi}(\theta) \doteq \mqty(e^{i\theta} & 0 \\ 0 & 1) \stackrel{\theta=\pi}{=} \mqty(-1 & 0 \\ 0 & 1),
\end{equation}
(which is just $(-1)\cdot\hat{\sigma}_{z}$) so that we get 
\begin{align}
\hat{\Phi}(\pi)\ket{\psi} &  \doteq \mqty(-1 & 0 \\ 0 & 1)\mqty(\frac{1}{\sqrt{2}} \\ \frac{1}{\sqrt{2}}) = \mqty(-\frac{1}{\sqrt{2}} \\ \frac{1}{\sqrt{2}}) \notag \\ & \doteq \frac{1}{\sqrt{2}}(\ket{\searrow} -\ket{\nearrow}) =: \ket{\psi'}.
\end{align}
But in case we insert a second beam splitter ($BS_{2}$ in the figure) at the point where the two trajectories cross, $\ket{\psi'}$ will change as 
\begin{equation}
\hat{U}_{H}\ket{\psi'} \doteq \frac{1}{\sqrt{2}}\mqty( 1 & 1 \\ 1 & -1)\mqty(-\frac{1}{\sqrt{2}} \\ \frac{1}{\sqrt{2}}) = \mqty(0 \\ -1) \doteq -\ket{\searrow}.
\end{equation}
So our simple qubit model predicts that we will always find a down moving photon in this setup, which has only picked up an unobservable phase of $\pi$ ($e^{i\pi} = -1$). 

Computing the probabilities for detecting an up- or downward traveling photon at the end of this setup gives, of course, $|-\braket{\nearrow}{\searrow}|^{2}=0$ and $|-\braket{\searrow}|^{2}=1$. The \emph{relative phase} between two kets is entirely responsible for the precise resulting behavior at $BS_{2}$;\footnote{\label{fn:phasdiff}Note that we have assumed both arms of the interferometer to be of equal length, so that none of the two states can pick up a phase due to spatial delay.} without the phase we would instead have 
\begin{equation}
\hat{U}_{H}\ket{\psi} \doteq \frac{1}{\sqrt{2}}\mqty( 1 & 1 \\ 1 & -1)\mqty(\frac{1}{\sqrt{2}} \\ \frac{1}{\sqrt{2}}) = \mqty(1 \\ 0) \doteq \ket{\nearrow},
\end{equation}
i.e.\! only photons moving \emph{up} at the end of the setup.

To check the predictions of this model one can use detectors ($d_{1}$ and $d_{2}$ in figure~\ref{fig:machzehnder1}) which give off a perceivable signal (a click if you will) upon incidence of a photon. With the phase shifter in place this (ideally) means only detections in $d_{2}$, and without the phase shifter (ideally) only in $d_{1}$, and experiments of this kind have of course been successfully performed \cite{aspect1986}.\footnote{In fact, varying the phase somewhat more than just $\theta\in\qty{0,\pi}$, one can appeal to probabilities of detection in either $d_{1}$ or $d_{2}$, where $\mathrm{Pr}_{x}^{\psi_{\theta}}(d_{1}) = |\braket{\nearrow}{\psi_{\theta}}|^{2} = \cos^{2}(\frac{\theta}{2})$ for $\ket{\psi_{\theta}} :=\frac{1}{2}\Big((1-e^{i\theta})\ket{\searrow} + (1+e^{i\theta})\ket{\nearrow}\Big)$, as results from the setup with a general phase shift. One can equally use a difference in path length, as mentioned in footnote~\ref{fn:phasdiff}, and this is what was done in \cite{aspect1986}, to confirm that the number of counts would conform to the predicted $\cos^{2}$-regularity.} 

Since the setup contains only one photon at a time, it seems surprising that it should matter to a photon traveling along the upper path whether there is a phase shifter in the lower one. But still, an analogous example can be constructed in the toy model by appeal to permutations instead of unitary matrices. The simplest type of permutation is a swap of two elements in an ordered sequence, and we will describe all permutations occurring in the example in terms of such swaps here. Thus, let $(jk)$ represent the swap of elements $j$ and $k$ in some ordered $n$-tuple ($n\geq j, n\geq k$). Then in the toy model we start out with $1\vee 2 \leftrightsquigarrow p_{\nearrow}$ as the epistemic state corresponding to the preparation of $\ket{\nearrow}(= \ket{0})$. The first beam splitter is represented by a permutation (23), which results in $1\vee 3$ (i.e.\! 3 will now be assigned the probability previously assigned to 2, which is $\frac{1}{2}$). The mirrors can be represented by (13), yielding $3\vee 1 = 1\vee 3$, so that not much happens here, just as in the QM treatment. In case the phase shifter is in, this can be modeled as a permutation corresponding to two successive swaps (12)(34) which then yield $2\vee 4$. And the second beam splitter will again correspond to (23), so that the final state is $3\vee 4$. But this distribution is the one corresponding to the quantum sate $\ket{1}=\ket{\searrow}$ so that the quantum predictions are indeed preserved. Equally, if the phase shifter is not inserted, this means that the permutation (12)(34) is left out, whence $1\vee 3$ will just be transformed into $1\vee 2$ at the second beam splitter, and we obtain the state that we started off with, again just as in QM.\footnote{All of these swaps can be implemented in the form of transformation matrices, as indicted in footnote \ref{fn:TransMat}. If one uses columns instead of rows to represent the epistemic states, (23) on $1\vee 2$, say, takes on the simple form ${\tiny \mqty(1 & 0 & 0 & 0 \\ 0 & 0 & 1 & 0 \\ 0 & 1 & 0 & 0 \\ 0 & 0 & 0 & 1)\mqty(1/2 \\ 1/2 \\ 0 \\ 0) = \mqty(1/2 \\ 0 \\ 1/2 \\ 0)}$.} 

Thus the toy model can indeed reproduce interference examples with the aid of resamplings of probability distributions. And resamplings can result in the toy-analogue of superpositions just as unitary transformations can result in quantum superpositions. We have here considered only a limited example with a certain fixed phase, but a mathematical generalization of Spekkens' work exists \cite{garner2013} which can handle arbitrary phase arguments in terms of probability vectors and transformation matrices. This achievement has lead several authors to conclude that ``a whole host of Mach-Zehnder interferometry experiments can be qualitatively reproduced by the theory[...].'' \cite[p.~79]{leifer2014}\footnote{Cf.\! also \cite[p.~3]{hardy2013} or \cite[p.~388]{fuchs2014a}.}

``But hold on!'', you may interject, ``How can a lack of knowledge account for the fact that what I do in the lower arm of the interferometer will influence \emph{all} photons in the setup, even if they take the upper route?'' And as well you should. We have here rather `blindly' applied the \emph{formal} tools of the toy model, which then appeared to nicely mirror some features of QM. But that permutations can be made to look like rotations on the Bloch sphere, and that these rotations in turn are homomorphic to unitary operations is a long shot from accepting that resamplings of a probability distribution (construed as a formal representation of a change in knowledge) can account for what goes on in a Mach-Zehnder interferometer. \emph{How} is it that our knowledge \emph{should} be affected by the putting in of the phase shifter? 

For all we know, many of the true states of systems in the setup---those representing something moving through the upper route---should \emph{not} be affected at all, whence, \emph{a fortiori}, neither should our knowledge of them. Building the example bottom-up, we would certainly not have guessed that putting in a phase shifter must result in interference, in case only one photon enters the setup. It is only our background knowledge of QM and the confirming experiments that allows us to concoct the toy model in the appropriate way, and it leaves us without any explanation as to \emph{why} our knowledge should change in this way. To be fair, we should here take into account that Spekkens aims to ``identify phenomena that are characteristic of states of incomplete knowledge regardless of what this knowledge is about.'' \cite[p.~2]{spekkens2007} A possible move at this point is thus to counter that we are informed by our experience with Mach-Zehnder interferometers, and that we should, in accord with the Calibration norm, adapt our epistemic states to the known frequencies in experiments with or without phase shifter. 

Still, there is some tension with the general philosophical stance of an EE view, as we have identified it in previous sections. A key motivation for an EE view is a certain preservation of common sense to secure a thorough basis for metaphysical realism. In particular, the introduction of true states $\lambda$ was identified to ensure microdefiniteness, and as such it raises hopes for finding a more complete physical theory that makes it possible (in principle) to give an account of how the world \emph{is} (beyond the QM description). Thus if the project is to serve its principal goal, it should at least \emph{allow} for an ontology of the true states that provides an explanation of the situation in question. 

Hence: What do $1,2,3,4$ represent, and how are they affected by the setup in such a way that the kind of probability update exemplified above is indicated? In fact, Spekkens and others seem to feel this need for explanation as well whence there \emph{is} a kind of (\emph{ex post}) explanation in the literature (cf.\! \cite{spekkens2008}; \cite{hardy2013}; \cite{leifer2014}). But we will only be able to suitably assess its plausibility after a discussion of two \emph{no-go theorems} below. 

We should now also look at \emph{combined} states of two (or more) simple systems in the model. Loosely following \cite[p.~11~ff.]{spekkens2007}, we can represent the simultaneous occurence of two true states $i,j\in\qty{1,\ldots,4}$ on two distinct systems $a,b$ respectively by a (symbolic) conjunction $i(a)\wedge j(b)$. Of course having such an epistemic state is prohibited by (\nameref{axm:KB}) since it would correspond to complete knowledge of the true states of both systems. But combinations of epistemic states, i.e.\! states of the form $[j(a)\vee k(a)]\wedge [\ell(b)\vee m(b)]$, with $j,k,\ell,m\in\Lambda$, and  $j\neq k, \ell\neq m$, are possible. These mimic simple \emph{product states} of QM, such as $\ket*{\psi^{(a)}}\ket*{\phi^{(b)}}$. 

A second possibility are states of the form $[j(a)\wedge k(b)]\vee [\ell(a)\wedge m(b)] \vee [n(a)\wedge o(b)]\vee [p(a)\wedge q(b)]$ with $j\neq \ell\neq n \neq p, \ \ k\neq m\neq o \neq q$. I.e.: it could be known, say, that both systems are in the same state, but not in \emph{which} state. Or it could be known that both are in different states, related by a certain specified permutation (transformation), but not which is in which. States of this form are supposed to mimic \emph{entangled states}, and \emph{prima facie} they do capture the essence of such states quite well.

To see this, take two systems which have been prepared in an entangled state, say $\ket{\pi} = \frac{1}{\sqrt{2}}(\ket{0,0} + \ket{1,1})$. Then this state implies that there is a probability of $1/2$ for each (sub)system to exhibit either of the two measurable values (0,1), but both systems are bound to exhibit the same value if the same observable is measured on them. Now consider a situation in which the two systems are separated spatially and two agents, $A$ and $B$ or `Alice' and `Bob', as they are usually called, perform measurements on them. Then at the very moment Alice measures `1', she will know that Bob will measure `1' as well, in case he measures the same observable. Phrased in terms of knowledge this is not so much of a surprise, but if we would endorse the orthodox interpretation instead, with its sudden change in the system's actual state due to the measurement, then Alice would be capable of `steering'\footnote{This is the much-used term introduced by Schr\"odinger \cite[p.~556]{schrodinger1935}.} Bob's system into some definite state, by choosing a certain kind of measurement to perform on \emph{her} system---and supposedly \emph{instantaneously} so at \emph{arbitrarily} large distances. 

From the point of view of the toy model this surprising consequence dissolves. Alice's state prior to measurement should be represented as $[1(a)\wedge 1(b)]\vee [2(a)\wedge 2(b)] \vee [3(a)\wedge 3(b)]\vee [4(a)\wedge 4(b)]$, since she knows that both systems are in the same state, even though she cannot know in which one. Accordingly, the measurement must result in something like $[1(a) \vee 2(a)] \wedge [1(b)\vee 2(b)]$, say. Treating the connectives in these symbolic formulae as actual conjunctions and disjunctions as in propositional logic for the moment, the latter statement straightforwardly follows from $[1(a)\wedge 1(b)]\vee [2(a)\wedge 2(b)]$ by case distinction and adding disjuncts. But the other, more important direction is not straightforwardly valid, since $[1(a) \vee 2(a)] \wedge [1(b)\vee 2(b)]$ is also true if $[1(a)\wedge 2(b)]$ holds, and taking into account that states pertaining to the same system mutually exclude each other,\footnote{It would hence be more appropriate to use exclusive disjunction $\dot{\vee}$ instead of $\vee$.} $[1(a)\wedge 1(b)]\vee [2(a)\wedge 2(b)]$ would actually be \emph{false}. With the epistemic state as given above however (the fourfold disjunction) and mutual state exclusion on the same system, Alice can draw the appropriate conclusion. 

Let us say that Alice chooses to measure $\qty{\qty{1, 3}, \qty{2, 4}}$ on her system and finds $1\vee 3$. Then she will come to know that \emph{both} systems must be in either of \emph{those} two true states (1 or 3). If she decides to measure $\qty{\qty{1, 2}, \qty{3, 4}}$ instead and finds $1 \vee 2$, then she comes to know that both systems must be in one of these states. So in fact performing both measurements in a row and obtaining these respective results Alice can come to the conclusion that both her \emph{and} Bob's system must have been in state 1 all along. So she instantaneously obtains information about the distant system. But since the act of measurement effects and unknown disturbance, the states of both systems may now (after both measurements) be different; the state of her system ($a$) could have changed to 2, in virtue of the disturbance effected by the second measurement. And assuming Bob performs the same protocol, he need not even obtain outcome $1\vee 2$ in the second measurement, since his system's state could have been changed to 3 in the first measurement and then $3 \vee 4$ would result in the second case. All that Alice can come to know is hence that during the \emph{first} measurement both systems \emph{were} in state 1; and hence this setup cannot be used as a means of \emph{communication}. This is of course reminiscent of the so called \emph{no-signaling theorems} in QM \citep[cf.~e.g.][pp.~393-394]{dickson2007}.

The truly crucial thing to realize, however, is that even if there is no real change in the \emph{true} state of $b$ due to Alice's measurement, it may still \emph{appear} this way in case one \emph{confuses} the epistemic state with the true state of the system. Since the model is supposed to provide evidence for an EE view of quantum states, the suggestion here seems to be that this sort of confusion is exactly what happens in orthodox QM. The example is indeed suggestive; \emph{prima facie} the $\psi$-epistemicist has a major advantage here. But the example is also \emph{selective}, and we all know (as do, of course, Spekkens and collaborators \cite[cf.][]{spekkens2007, harriganspekkens2010, bartlett2012}) that the existence of Bell-type inequalities raises doubts about this kind of an interpretation of entanglement. This is an issue we will reassess in the discussion. 

For completeness' sake, it should not go unmentioned that Bartlett et al. \cite{bartlett2012} have worked out a model similar in spirit to Spekkens' original toy model, which reproduces a bunch of phenomena in continuous-range systems.\footnote{It is not clear that the model fits into the OM approach or whether it can be made to do so. This does not pose a problem for us though, since we are concerned more generally with EE views of quantum states.} A thorough discussion of this model exceeds the scope of this paper, whence we only give a brief review. In short, the authors show that putting an epistemic restriction (similar to (\nameref{axm:KB})) on \emph{Liouville mechanics}, the statistical version of classical Hamiltonian mechanics, one obtains a theory which is ``operationally equivalent'' (p.~2) with a subtheory of QM, which they spell out to mean that 

\begin{quote}there is a one-to-one mapping between the preparations, measurements, and transformations that are allowed in the first theory and those that are allowed in the second and [that] the statistics predicted for every possible experiment in the first theory are precisely the same as those predicted for the corresponding experiment in the second theory. (p.~15)\end{quote}

Because the model is a restricted version of classical statistical mechanics, the true states of systems in question are points $\bm{z} = (q_{1},\ldots, q_{3n}, p_{1}, \ldots, p_{3n})$ in phase space (for $n$ mass points with 3 position and 3 momentum coordinates $q_{i}, p_{j}$; we set $6n \equiv N$). 

The epistemic restriction is twofold. First of all, Bartlett et al. define the set 
\begin{equation*}
L_{+}(\Gamma) :=\qty{\mu \vert \mu : \Gamma \rightarrow \mathbb{R}, \mu\geq 0, \int_{\Gamma}\mu(\bm{z})\dd[N]{\bm{z}}=1}
\end{equation*}
of (Liouville) probability densities on phase space $\Gamma$. Then for any $\mu$ to be considered a \emph{valid} distribution for their model, it is required that (i) the covariance matrix $\gamma(\mu)$ satisfies the `classical uncertainty principle' $\gamma(\mu) + i\lambdaslash\Sigma \geq 0$, where $\lambdaslash$ is a free parameter and
\begin{equation*}
\Sigma := \tiny{\mqty (0 & -1 & 0 & 0 & \ldots \\ 1 & 0 & 0 & 0 & \\ 0 & 0 & 0 & - 1 & \\  0 & 0 & 1 & 0 & \\ \vdots & & & & \ddots )},
\end{equation*}
and that (ii) $\mu$ has maximum entropy 
\begin{equation*}
S(\mu) = -\int_{\Gamma} \mu(\bm{z})\ln(\mu(\bm{z}))\dd[N]{\bm{z}}
\end{equation*}
over $\Gamma$ among all phase space distributions with the same covariance matrix \cite[p.~5]{bartlett2012}. The covariance matrix of a distribution that depends on multiple coordinates $z_{i}, z_{j}$ (in phase space, in this case) describes, in components $\gamma_{ij}$, (twice) the covariance $\ev{\Big(z_{i}-\ev{z_{i}}\Big)\Big(z_{j}-\ev*{z_{j}}\Big)}_{\mu}$, i.e. the correlation of departures from the mean values $\ev{z_{i}}_{\mu}, \ev*{z_{j}}_{\mu}$ according to $\mu$ \cite[cf.][p.~361]{janes2003}. The bite of (i) is that it parallels an actual formulation of the uncertainty relations, and thus ensures that in the restricted Liouville mechanics, relations such as $\Delta p_{x}\Delta x\geq \lambdaslash /2$ hold (for adjustable $\lambdaslash$). (ii) on the other hand ``ensures that an agent should have the maximum uncertainty about the physical state of the system consistent with knowing the means and the covariance matrix.'' \cite[p.~5]{bartlett2012} The valid distributions satisfying (i) and (ii) are all of Gaussian form.  

The theory which results is thus operationally equivalent (in their sense) to what they call ``Gaussian quantum mechanics'' (p.~2), the part of QM ``including only those preparations, measurements, and transformations that have Gaussian Wigner representations [...].'' (ibid.) 
However, we shall argue below that even in this more elaborate model, some quantum phenomena---and arguably the most important ones---\emph{cannot} be reproduced. 

\section{The Impact of No-Go Results}
The EE view in the form of $\psi$-epistemic models has been confronted with a bunch of no-go results, the most influential one being that of Matthew Pusey, Jonathan Barrett, and Terry Rudolph (PBR), published in \emph{Nature} in 2012. In a preprint-version of their paper, the authors also proposed an error-tolerant version of the experimental conditions described in the proof, allowing for an actual test of the diverging predictions, which has indeed been successfully implemented shortly after \citep{nigg2012}. The theorem is supposed to demonstrate that a $\psi$-epistemic model for QM is not feasible, but its proof is of course not free of assumptions, and we need to dedicate some careful attention to these. The theorem is first demonstrated for quantum states with overlap $\braket{\phi}{\psi} = \frac{1}{\sqrt{2}}$ and then generalized to states with arbitrary overlaps. We shall restrict ourselves to a discussion of the former case and only briefly sketch how the generalization is established.

We shall see, however, that this particular theorem may not be as devastating to an EE view as its popularity suggests. Its main impact rather is that it has spawned off a new level of the debate and inspired a host of further no-go results. We will subsequently review and analyze a theorem by L. Hardy \cite{hardy2013}, which may, effectively, have deeper implications for current successes of facilitating an EE view than does the PBR theorem.

\subsection{The PBR Theorem}\label{sec:PBR}
To show the incompatibility of QM with $\psi$-epistemic models, PBR consider two qubit systems which are supposed to be prepared entirely \emph{independently} of one another, each in one of the two quantum states $\ket{0}$ and $\ket{+}$. Note that these states are \emph{non-orthogonal}, whence, in line with the discussion above, they are plausible candidates for P-states with overlapping associated probability distributions (signifying $\psi$-epistemicity). 

The first thing to realize\footnote{Note, however, that we here provide an analysis that dissents from the original presentation in \cite{pbr2012}, and in part draws on other analyses (where referenced).} is that the assumption of independence translates into two different formal requirements in the two different formalisms (QM proper and the OM approach), whose intertranslation requires a bridging assumption. In QM, independence can be represented by the use of product states; thus, if $\ket{\Psi}$ denotes the total P-state of the two systems, we can translate the assumption of preparation independence into
\begin{equation}\label{eq:prod1}
\tag{Prod. 1} 
\ket{\Psi}\in\lbrace \underbrace{\ket{0}\ket{0}}_{=:\ket{\Psi_{1}}}, \underbrace{\ket{0}\ket{+}}_{=:\ket{\Psi_{2}}}, \underbrace{\ket{+}\ket{0}}_{=:\ket{\Psi_{3}}}, \underbrace{\ket{+}\ket{+}}_{=:\ket{\Psi_{4}}}\rbrace =: \mathcal{P}
\end{equation}
(with $\ i,j\in\qty{0, +}$, $\mathcal{P}$ for `preparation'). Now in the OM-approach two true states $\lambda_{1}$ and $\lambda_{2}$ need to be specified, since two systems are concerned. The next non-trivial assumption (the bridging principle) is that the state space is \emph{separable} in an appropriate manner, i.e.: 
\begin{equation}\label{eq:Sep}
\tag{Sep.} 
 \Lambda_{\Psi} = \Lambda_{1}\times\Lambda_{2}, 
\end{equation}
with $\Lambda_{1}$ and $\Lambda_{2}$ the state spaces of the two systems respectively \cite[cf.][]{spekkens2012, leifer2014}. This separability assumption amounts to assuming that, ``when modeling independent local preparations, there are no additional properties of the joint system that are not derived from the properties of the individual systems.'' \citep[p.~100]{leifer2014} It is hence basically the ontological assumption which justifies the next step. Namely, given~(\ref{eq:Sep}), the independence-assumption can be translated into a classical probabilistic language, suitable for the OM approach, as
\begin{equation}\label{eq:prod2}
\tag{Prod. 2} 
p_{j}(\lambda_{1}, \lambda_{2}) = p_{k}(\lambda_{1})p_{\ell}(\lambda_{2}), \ \ j\in\qty{1,\ldots,4}, \ \ k,\ell\in\qty{0,+}
\end{equation}
\citep[cf.][p.~477]{pbr2012}, where $p_{j}(\lambda_{1}, \lambda_{2}) := p(\lambda_{1}, \lambda_{2}|\Psi_{j})$ and \linebreak $p_{k}(\lambda) := p(\lambda|\psi_{k})$  $(j\in\qty{1,\ldots,4},k\in\qty{0,+})$. The two conditions (\ref{eq:prod1}) and (\ref{eq:prod2}) are neither logically equivalent, nor does (\ref{eq:prod1}) straightforwardly imply \linebreak (\ref{eq:prod2}). But arguably (\ref{eq:prod1}) \emph{conceptually} implies (\ref{eq:Sep}), and (\ref{eq:Sep}) conceptually implies (\ref{eq:prod2}): If we can prepare two systems in (sufficient) isolation from one another, we use a tensor product in QM to represent the (P-)state of a composite system. But if we use such a product state, we assume both component systems to \emph{be} (sufficiently) independent of one another. And given that we hence assume their respective (\emph{true}) states to also be independent of one another, i.e., given (\ref{eq:Sep}) we would model this very situation by a mere product-distribution in a `traditional' probabilistic setting.\footnote{This of course means that $p_{j}(\lambda_{1}|\lambda_{2}, \psi_{k}^{(2)})=p_{j}(\lambda_{1}), \ j,k\in\qty{0,+}$, with $\psi_{k}^{(2)}$ the P-state for the \emph{second} system, and analogously for $p_{k}(\lambda_{2})$.} Hence it fully suffices to claim that (\ref{eq:prod1})$\rightarrow$(\ref{eq:Sep}), and that (\ref{eq:Sep})$\rightarrow$(\ref{eq:prod2}) to get the central premise:
\begin{equation}\label{eq:PIndep}
\tag{P.-Indep.} 
\text{(Prod. 1)}\rightarrow\text{(Prod. 2)}
\end{equation}

Suppose now \cite{pbr2012} that there is a $\Delta$ such that $\lambda_{1}, \lambda_{2}\in\Delta$, i.e.\! there are true states on both of the two systems which lie in the overlap-region for probability distributions possibly associated with distinct quantum states $\ket{0}$ and $\ket{+}$. Basically: assume $\psi$\emph{-epistemicity} to hold, as defined in section~\ref{sec:OntMod}, but for true states of two different systems. Also, fix some lower limit $q>0$ such that $p_{k}(\lambda_{1})\geq q, p_{\ell}(\lambda_{2})\geq q$ for $k,\ell\in\qty{0,+}$ and $\lambda_{1}, \lambda_{2}\in\Delta$. Then by (\ref{eq:PIndep}), we get that
\begin{equation}\label{eq:Delta}
\tag{$\Delta$} 
p_{\Psi}(\lambda_{1}, \lambda_{2}) \geq q^{2}, \ \ \  \forall\Psi\in\mathcal{P} \ \forall \lambda_{1},\lambda_{2}\in\Delta  
\end{equation}

	\begin{figure}
	\centering
	\includegraphics[scale=0.55]{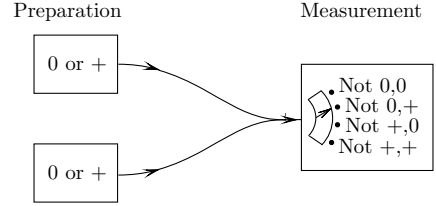}
	\caption[The PBR setup]{\small{Each system is prepared in one of two quantum states; the entangled measurement performed on both systems simultaneously then finds out which of the four possible product states was not prepared. (Cf. \cite{pbr2012} for a similar illustration).}}
	\label{fig:PBRSet}
	\end{figure}
We call this intermediate result `$(\Delta)$' because the existence of some $\Delta$ (i.e.\! the positivity of the product distribution $p_{\Psi}$ on some set of non-zero measure) regardless of the specific preparation on each system, is crucial. It is also crucial to realize that the preparation procedures on both systems do the same thing, i.e., prepare either $\ket{+}$ or $\ket{0}$, whence the (total) range of true states $\lambda$ possibly resulting from the preparations is identical for the two systems. This (in concert with (\ref{eq:PIndep})) justifies why it even makes sense to consider this setup for \emph{two systems} as a means to check for the possibility of a $\psi$-epistemic model, where the assumption of an overlap was previously formulated w.r.t.\! to the states of \emph{one and the same} system.

Now the measurement executed on the two systems is performed by bringing them together in one measurement-device and measuring them jointly (cf.\! figure~\ref{fig:PBRSet}). A measurement of this kind is called \emph{global}, since all the systems in some total state $\ket{\Phi}$ are measured together, and only information about their total state $\ket{\Phi}$ is acquired. Among such global measurements, one can further distinguish measurements which have only product states as possible outcomes from such which have at least one entangled state among their outcomes. That is to say, in the latter case, the operators used to describe the measurement have entangled eigenvectors. These are then (unsurprisingly) called \emph{entangled measurements} \citep[cf.~e.g.][pp.~219-220]{wootters2006}. 

The measurement considered by PBR is exactly such an entangled (global) measurement. Furthermore, it is projective, resulting in an M-state out of the following set:
\begin{align*}
\mathcal{R} := \Big\{ & \ket{\phi_{1}} = \frac{1}{\sqrt{2}}(\ket{0}\ket{1} + \ket{1}\ket{0}), \\ & \ket{\phi_{2}} = \frac{1}{\sqrt{2}}(\ket{0}\ket{-} + \ket{1}\ket{+}),\\
& \ket{\phi_{3}} = \frac{1}{\sqrt{2}}(\ket{+}\ket{1} + \ket{-}\ket{0}), \\ &\ket{\phi_{4}} = \frac{1}{\sqrt{2}}(\ket{+}\ket{-} + \ket{-}\ket{+})\Big\}
\end{align*}
\citep[cf.][p.~476]{pbr2012}.\footnote{We call this set `$\mathcal{R}$' for `result', and for notational simplicity we will later also use this letter to refer to the measurement (POVM) associated with the outcome states in $\mathcal{R}$.} What we now see is that for each of the $\ket*{\phi_{j}}\in\mathcal{R}$ there is a $\ket{\Psi_k}\in\mathcal{P}$ which is \emph{orthogonal} to it (whence the global property to be measured is which of the states was \emph{not} prepared; cf. figure~\ref{fig:PBRSet}). For instance,
\begin{eqnarray}\label{eq:probzero}
 \braket{\phi_{1}}{\Psi_{1}} & = &\frac{1}{\sqrt{2}}(\bra{0}\!\otimes\!\bra{1} + \bra{1}\!\otimes\!\bra{0}) \ket{0}\!\otimes\!\ket{0}=\notag \\ &=& \frac{1}{\sqrt{2}}(\braket{0} \braket{1}{0} + \braket{1}{0}\braket{0}) = \notag \\
 & = &\frac{1}{\sqrt{2}}(1\cdot 0 + 0 \cdot 1) = 0,
\end{eqnarray}
and (because of the way we have indexed the states) in general $\braket{\phi_{j}}{\Psi_{j}}=0$.

But recall that the connection between the Born probabilities and the probability distributions in the OM was established by an integral over the epistemic state and the response function (formula~(\ref{eq:OMProb})). This integral must now take the form
\begin{equation*}
\mathrm{Pr}_{\mathcal{R}}^{\ket{\Psi_{j}}}(k) = \int \dd{\lambda_{1}}\int \dd{\lambda_{2}}p_{j}(\lambda_{1}, \lambda_{2})\xi_{\mathcal{R}}^{\phi_{k}}(\lambda_{1}, \lambda_{2})
\end{equation*}
\citep[cf.][p.~477]{pbr2012}, with $\xi_{\mathcal{R}}^{\phi_{k}}(\lambda_{1}, \lambda_{2})$ the response function for outcome $k$. 

Moreover, it is plausible to require that
\begin{equation}\label{eq:outc}
\tag{Outc.} 
\sum_{k=1}^{4}\xi_{\mathcal{R}}^{\phi_{k}}(\lambda_{1}, \lambda_{2})= 1, \ \ \ \forall (\lambda_{1}, \lambda_{2})\in\Lambda_{\Psi},
\end{equation}
i.e.\! that there will always be \emph{some} outcome for \emph{all} the states that may result from the preparation \citep[cf.~e.g.][p.~1]{aaronson2013}. Of course this is quite an idealization, and we may assume that (\ref{eq:outc}) is only required to hold up to expected experimental noise and error. 

Since $p_{j}(\lambda_{1}, \lambda_{2})$ is at least $q^{2}$ on a set $\Delta$ of non-zero measure in virtue of ($\Delta$), it must hold that 
\begin{align}\label{eq:PBRContr}
\exists k \forall j: \mathrm{Pr}_{\mathcal{R}}^{\ket{\Psi_{j}}}(k)& = \int \dd{\lambda_{1}}\int \dd{\lambda_{2}}p_{j}(\lambda_{1}, \lambda_{2})\xi_{\mathcal{R}}^{\phi_{k}}(\lambda_{1}, \lambda_{2})  > 0 \notag \\ & \stackrel{!}{=} \qty|\braket{\phi_{k}}{\Psi_{j}}|^{2} = 0 \text{~~~for } j=k ~~~\lightning \tag{PBR}
\end{align}
(with $j, k \in\qty{1,\ldots,4}$). This is the PBR contradiction. (\ref{eq:prod1}), (\ref{eq:PIndep}), and (\ref{eq:outc}) taken together with the definition of $\psi$-epistemicity and the general assumptions of the OM framework (short: $\qty{\text{OM}}$), lead to a contradiction; hence PBR conclude: 
\begin{equation}
\qty{\text{OM}}, \text{(\ref{eq:prod1}), (\ref{eq:PIndep}), (\ref{eq:outc})} \vdash \neg(\psi\text{-epistemicity})
\end{equation}
Expressed differently, this means that any $\psi$-epistemic OM cannot maintain (\ref{eq:prod1}), (\ref{eq:PIndep}), and (\ref{eq:outc}) together, all of which are \emph{prima facie} reasonable assumptions. So possibly there are no suitable $\psi$-epistemic OMs.

We have restricted our attention to the two-system case, but the result of PBR is generalized \cite[p.~476~ff.]{pbr2012} using tensor-product states $\ket{\Psi} = \ket{\psi_{1}}\otimes\ldots\otimes\ket{\psi_{n}}$ of arbitrary finite cardinality $n$, where each system is prepared in either $\ket{0}$ or $\ket{+}$ ($\psi_{j}\in\qty{0,+}, \ \forall 1\leq j\leq n$). This allows for states with an overlap different from that between $\ket{0}$ and $\ket{+}$ to be used in the preparation. 

But how deep is the impact of PBR's result really? Should it be taken to rule out $\psi$-epistemic OMs \emph{tout court}? To answer this question, we should look at each of the premises of the proof separately.

A lot of criticism towards the premises of the PBR theorem can be found especially in an article by Schlosshauer and Fine \cite{schlosshauerfine2012}. Notably, they first of all refrain from even using the terminology of `$\psi$-epistemic' and `$\psi$-onitc' models, and refer to these classes of models as `mixed' and `segregated' instead (to them this terminology is ``less charged'' \cite[p.~4]{schlosshauerfine2012}). Thus, the general aptness of the very \emph{definition} of a $\psi$-epistemic model used in the OM approach may of course be put into question (and hence the premise $\qty{\text{OM}}$). A whole other set of criteria for understanding the wave function as a representation of knowledge (in the EE sense) may of course be available. Schlossauer and Fine then also show a way of transforming mixed models into segregated ones and \emph{vice versa}, thus lessening the appeal of the definitions from section~\ref{sec:OntMod} as indeed reflecting a distinction between something that represents knowledge and something that represents something real.\footnote{These charges of transformability between the two types of models are, however, challenged by Leifer \cite[p.~113-114]{leifer2014}.} 

Beyond that, Schlosshauer and Fine suggest to augment the spectrum of outcome values associated with the measurement with so called `no-shows', i.e.\! to allow for measurements with no discernible outcome at all, and hence to modify the connection between the Born probabilities and the probability distributions in the OM-framework accordingly. One crucial step of PBR's theorem is to require (\ref{eq:outc}) and (\ref{eq:outc}) is, as we noted, somewhat idealized. Modifying this requirement in such a way that, given that the true state is in the overlap region, there will be a probability of obtaining no outcome at all, determined \emph{by the true state itself}, obviously blocks the inference to $\neg(\psi\text{-epistemicity})$. Schlosshauer and Fine refer to this as a ``built-in inefficiency'' \cite[p.~2]{schlosshauerfine2012}, since the assumption is that there is something about the measured system itself which lets the probability of a (discernible) outcome drop in the appropriate region. 

A bit more precisely, the general recipe goes like this: Determine some probability $\xi^{\emptyset}_{\mathcal{R}}(\lambda_{1},\lambda_{2})$ of getting a \emph{null-outcome} $\emptyset$ (i.e. something that cannot be recognized properly as an outcome on the measuring device), sufficiently high for the $\lambda\in\Delta$, so that the QM statistics is reproduced, but now from probabilities \emph{conditional} on the fact that a discernible outcome was measured at all (i.e. by postselecting the statistics for runs in which there was a determinate outcome). Then for the set of outcomes $\qty{\phi_{1},\ldots\phi_{4}, \emptyset}$, the resulting version of (\ref{eq:outc}) is \emph{not} violated and no contradiction arises. Under these assumptions, all that the PBR-result shows is ``how inefficiencies arise as a fundamental property of certain hidden-variables models [...].'' \cite[p.~2]{schlosshauerfine2012} 

This is a kind of `prism model', which the reader may be familiar with from the context of Bell inequalities. However, there is a certain \emph{ad hoc}-ness to assuming that the true states from the overlap mysteriously sabotage the measurement procedure just to recover the quantum statistics. Thus we may be inclined to put more doubt on the justifyability of Schlosshauer and Fine's no-show assessment than on PBR's own one.

The various assumptions underlying (\ref{eq:PIndep}) are also under scrutiny in Schlosshauer and Fine's article. They think that ``[c]orrelations [...] cannot be ruled out, even if the preparations appear to be independent, because procedures for preparing the individual subsystems may occur together closely in spacetime or share common sources of energy, as well as a common past.'' \citep[p.~3]{schlosshauerfine2012} In our reconstruction we may take this criticism to aim at the validity of the implication (\ref{eq:prod1})$\rightarrow$(\ref{eq:Sep}), and so indirectly at the validity of (\ref{eq:PIndep}). But (\ref{eq:Sep}) can be weakened to the condition (call it `(Sep.$^{*}$)') that, if there is a $\lambda$ in the support of \emph{each} of the epistemic states associated with the multiple systems and respective quantum states, then there is also some $\lambda_{c}$ in the support of the \emph{common} distribution $p_{\Psi}$ associated with the product state $\ket{\Psi} = \ket{\psi_{1}}\otimes\ldots\otimes\ket{\psi_{n}}$ (ibid.). The exact nature (and structure) of $\lambda_{c}$ can then be left completely unspecified. From this one neither gets the condition (\ref{eq:PIndep}), because (Sep.$^{*}$) does not imply (\ref{eq:prod2}), but rather that $p_{\Psi}(\lambda_{c})>0$ (call this `(Pos.)'). Nor does one get the (exact) $q^{2}$-result (\ref{eq:Delta}), which follows from (\ref{eq:prod2}), not (Pos.). But since the weaker (Pos.) is obviously sufficient to derive a contradiction (i.e.\! (\ref{eq:outc}) would still have to be violated) it appears that PBR's conclusion $\neg(\psi\text{-epistemicity})$ is not really warranted, and that the theorem need not be considered as applying to $\psi$-epistemic models after all. But this move of Schlosshauer and Fine is only possible on the pains of replacing (\ref{eq:Sep}) by (Sep.$^{*}$) and hence by denying (\ref{eq:PIndep}), or in other words: by assuming that the systems in question \emph{cannot} be prepared (sufficiently) independently of one another.

Additionally,  Schlosshauer and Fine criticize that PBR \emph{implicitly} assume that the response functions $\xi_\mathcal{R}^{\phi_{j}}(\lambda_{1}, \lambda_{2})$ do not depend on $\Psi$, and they propose \cite[p.~2~ff.]{schlosshauerfine2012} that models which avoid the problem raised by PBR can be constructed, in case this assumption is dropped. They call the class of models presupposed by PBR \emph{state-independent}. Leifer \cite[p.~111]{leifer2014}, in contrast, thinks that ``this criticism is simply a misunderstanding of what is meant by the term `ontic state' in the ontological models framework'', and goes on to demonstrate an example of how models can \emph{trivially} reproduce the Born probabilities in case state dependence is allowed (that is, in case $\xi$ is also conditional on the prepared quantum state $\Psi$). In a similar vein, Ballentine refers to such models as ``functionally $\psi$-ontic'', because 

\begin{quote}[t]he most important structure of the model is the separation of \emph{preparation} from \emph{measurement}, with information passing only via the ontic state variables. If the state $\psi$ has a direct effect on the measurement outcome, then $\psi$ should be classified as an \emph{ontic} variable. \cite[p.~6; emphasis in original]{ballentine2014}\end{quote} 

Hence the assumption of state independence may be considered as justified (or -fiable); the introduction of state dependence \emph{conceptually undermines} the very idea behind $\psi$-epistemicity in the OM-approach.

Nevertheless, Schlosshauer and Fine's conclusions on the impact of the PBR theorem remain de-emphasizing:
\begin{quote}PBR show that state-independent models of composites formed using systems with mixed [$\psi$-epistemic -- FB] models face restrictions. It is vital to see that those restrictions do not imply any difficulty for models of the \emph{components} themselves. The PBR theorem is not a no-go theorem for the component systems[...]. \cite[p.~4~ff.; my emphasis -- FB]{schlosshauerfine2012}\end{quote} 

And indeed, the theorem is \emph{not} concerned with several quantum states of a \emph{single} system, but only has an impact on overlapping epistemic states via the detour of using product states of compound systems. One may jump in at multiple points and criticize the assumptions that bridge the gap, as we have just seen. Moreover, Lewis et al. \cite{lewisjennings2012} actually have provided two variants of a $\psi$-epistemic model which become possible in case (\ref{eq:PIndep}) is dropped.\footnote{They do not use our formal reconstruction of preparation independence in their article though, but instead give the informal characterization that ``situations where quantum theory assigns independent product states are presumed to be completely describable by independently combining the two purportedly deeper descriptions for each system.'' \cite[p.~1]{lewisjennings2012}} But the models are utterly formal, and they also concede: 

\begin{quote} None of these models is intuitive or motivated by physical principles or considerations. The primary motivation for exploring the possibility of $\psi$-epistemic models is to understand the formal limitations of reproducing quantum theory from a deeper theory. \citep[p.~4]{lewisjennings2012}\end{quote} 

Their conclusion w.r.t.\! the latter aim is that ``any similar no-go theorem will also require nontrivial assumptions beyond those required for a well-formed ontological model.'' \citep[p.~1]{lewisjennings2012} We can take from this that, while restricting the possibility of $\psi$-epistemic models, the PBR theorem (and similar results) should not count as a \emph{full} no-go theorem for these models, in the sense of demonstrating their \emph{impossibility}. They all rely on additional assumptions and can hence maximally limit the attractiveness of $\psi$-epistemic hidden variable models, or more precisely, show their incompatibility with these very assumptions.

Regarding the existence of other such theorems, the PBR paper has indeed caused a whole landslide of publications which put forward theorems purportedly showing the impossibility of $\psi$-epistemic models (so, in fact, their incompatibility with other plausible assumptions).\footnote{A partial survey of the developments up to the year 2014 can be found in \cite{leifer2014}.} One such theorem that we should now take a closer look at is that of Hardy \cite{hardy2013}. This will give us a chance to directly confront some of the purported achievements of Spekkens' toy model. 

\subsection{Hardy's Theorem}
The gist of Hardy's theorem can best be captured by appeal to an interferometry example like the ones we had met with in section~\ref{sec:toymod}. 

\begin{figure}[h]
	\centering
	\includegraphics[scale=0.4]{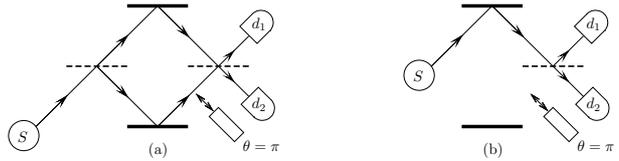}
	\caption[Hardy's theorem]{\small{(a) is the Mach-Zehnder setup as discussed in section~\ref{sec:toymod}. In (b) the photon is emitted somewhere along the upper trajectory, whence the phase shifter in the lower trajectory should have no effect.}}
	\label{fig:machzehnder2}
	\end{figure}
In Hardy's own words, the argument based on the following example amounts to a ``version of the popular argument for something going both ways[...].'' \cite[p.~6]{hardy2013} Consider, in contrast to the Mach-Zehnder example we have discussed in section~\ref{sec:toymod}, an altered setup where the source of photons is placed somewhere along the upper route (cf.\! figure~\ref{fig:machzehnder2}~(b)). In this altered setup, it should not matter whether the beam splitter is inserted or not; whereas in the original Mach-Zehnder example we would obtain either $-\ket{\searrow}$ or $\ket{\nearrow}$ at the end of the interferometer, depending on whether the phase shifter was in or not, we will here simply have 
\begin{equation}
\hat{U}_{H}\hat{\sigma}_{x}\ket{\nearrow} = \hat{U}_{H}\ket{\searrow} = \frac{1}{\sqrt{2}}(\ket{\nearrow} - \ket{\searrow}),
\end{equation}
whence detection at $d_{1}$ and $d_{2}$ will be equiprobable. 

Now consider the state $\ket{\psi}=\frac{1}{\sqrt{2}}(\ket{\nearrow} + \ket{\searrow})$ as prepared by the first beam splitter in the setup of figure~\ref{fig:machzehnder2}~(a), and the state $\ket{\phi}=\ket{\nearrow}$ as prepared by the source in the setup of figure~\ref{fig:machzehnder2}~(b). These two states are non-orthogonal and hence could well be taken to have overlapping supports in a $\psi$-epistemic model. In this context, we can understand this claim such that it is not impossible for the first beam splitter to prepare a photon which is \emph{actually} traveling up, and that $\ket{\psi}$ is again just indicative of our lack of knowledge about the true state, i.e. the true path the photon takes. 

But then it should make no difference for the photons actually traveling up whether the phase shifter is inserted in the lower path or not. Thus, denote the full set of true states associated with $\ket{\phi}$ by $\Lambda_{\ket{\phi}}$, and the subset of those resulting in a click in detector $d_{1}$ or detector $d_{2}$ by $\Lambda_{\ket{\phi}}^{d_{1}}$ and $\Lambda_{\ket{\phi}}^{d_{2}}$ respectively. One can also associate a given setting of the phase shifter (in or out) to these sets, which we indicate (following \cite[p.~6]{hardy2013}) by the notation $\Lambda_{\ket{\phi}}^{d_{j}}[\theta]$ ($j\in\qty{1,2}, \theta\in\qty{0,\pi}$). But since the choice of $\theta$ as $0$ or $\pi$ should not alter the behavior of the photon going along the upper path, we obtain a kind of \emph{invariance}:
\begin{equation}\label{eq:thetainvar}\tag{INVAR}
\Lambda_{\ket{\phi}}^{d_{j}} = \Lambda_{\ket{\phi}}^{d_{j}}[\theta = 0] = \Lambda_{\ket{\phi}}^{d_{j}}[\theta = \pi], \ \ \ j\in\qty{1,2}.
\end{equation}
Assume that the photon is bound to end up in one of the detectors---neglecting, of course, experimental errors, i.e., photons getting absorbed somewhere along the way or detectors not firing upon incidence---whence it should hold that 
\begin{equation}\label{eq:union}\tag{TOT}
\Lambda_{\ket{\phi}} = \Lambda_{\ket{\phi}}^{d_{1}}\cup\Lambda_{\ket{\phi}}^{d_{2}},
\end{equation}
irrespective of the choice of $\theta$. Now consider the set of true states $\Lambda_{\ket{\psi}}$ associated with $\ket{\psi}$ (the state prepared by the first beam splitter). We had established above that in case the phase shifter is in ($\theta = \pi$), the state $\ket{\psi}$ will not result in any clicks in detector $d_{1}$. Thus it should hold that 
\begin{eqnarray}\label{eq:disjoint1}
\Lambda_{\ket{\psi}} \cap \Lambda_{\ket{\phi}}^{d_{1}}[\theta = \pi] & = & \varnothing \notag \\
\Leftrightarrow \Lambda_{\ket{\psi}} \cap \Lambda_{\ket{\phi}}^{d_{1}} & = & \varnothing,
\end{eqnarray}
where the equivalence follows from \eqref{eq:thetainvar}. Analogously, in case the phase shifter is out ($\theta = 0$), there will be no clicks in detector $d_{2}$ if $\ket{\psi}$ is prepared, so that 
\begin{eqnarray}\label{eq:disjoint2}
\Lambda_{\ket{\psi}} \cap \Lambda_{\ket{\phi}}^{d_{2}}[\theta = 0] & = & \varnothing \notag \\
\Leftrightarrow \Lambda_{\ket{\psi}} \cap \Lambda_{\ket{\phi}}^{d_{2}} & = & \varnothing.
\end{eqnarray}
But from (\ref{eq:union}), (\ref{eq:disjoint1}), and (\ref{eq:disjoint2}) it now follows that $\Lambda_{\ket{\psi}} \cap \Lambda_{\ket{\phi}} = \varnothing$, whence there is no intersection in the sets of true states associated with the two non-orthogonal states $\ket{\psi}$ and $\ket{\phi}$. This in turn means that the epistemic states for the two preparation methods associated with $\ket{\psi}$ and $\ket{\phi}$ cannot have overlapping supports. Thus, it seems, this situation \emph{cannot} be understood $\psi$-epistemically \cite[p.~7-9]{hardy2013}.

Of course this is not yet a no-go theorem for $\psi$-epistemic OMs but merely an example. In the remainder of his paper, Hardy provides a generalization, first for finite Hilbert spaces, for which it is shown that non-orthogonal states with a certain lower bound quantum probability $|\braket{\phi}{\psi}|^{2}$ (which depends on the dimension of the Hilbert space) will result in distributions with non-overlapping supports (pp.~9-13). For an infinite dimensional Hilbert space, the result is then shown to hold regardless of the quantum probability (p.~12). For a rigorous, general proof one of course needs to abstract from beam splitters, mirrors, and phase shifters. The phase shifter, for instance, is replaced by a general unitary transformation with some general parameter $m$ to be varied (instead of the phase shift $\theta$) \citep[p.~10~ff.]{hardy2013}. 

But of course, a few crucial assumptions also have to be made to run this proof, just as in the PBR case. For the proof of Hardy's theorem the following two principles have to be assumed \citep[cf.][pp.~4-5]{hardy2013}:  
\begin{PC}[PC]\label{dfn:PC}
The ontic state, $\lambda$, is sufficient to determine whether any outcome of any measurement has probability equal to zero of occurring or not.
\end{PC}
\begin{ROD}[ROD]\label{dfn:ROD}
Any quantum transformation on a system which leaves a particular given pure quantum state, $\ket{0}$, unchanged can be implemented in such a way that it does not affect the underlying ontic states, $\lambda\in\Lambda_{\ket{0}}$, in the ontic support of $\ket{0}$.
\end{ROD}

Note that Hardy first assumes a stronger principle of ontic indifference, which is supposed to hold for any arbitrary quantum state $\ket{\psi}$ instead of a particular one ($\ket{0}$). He then demonstrates that the weaker principle (\nameref{dfn:ROD}) is sufficient to run the proof \citep[p.~12]{hardy2013}. The `ontic support' is of course the support of the epistemic state, i.e., the set of true states $\lambda$ which may result from the preparation procedure associated with $\ket{\psi}$. (\nameref{dfn:PC}) is also a rather weak principle, since the true state only determines whether an outcome has probability zero or not, instead of determining the exact probability.\footnote{The intuition behind the use of `possibility' in the name of (\nameref{dfn:PC}) is certainly that the true state determines whether an outcome is possible at all. But this is obviously not correct, as probability zero is not synonymous with something being impossible on all accounts of probability. The limit frequency of an event in an infinite random sequence may be zero even though this event is not impossible.}

We have seen both of these principles at work in the example considered above. (\nameref{dfn:PC}) is used to define the sets of states which may give rise to a detection by $d_{1}$ or $d_{2}$ respectively. (\nameref{dfn:ROD}) is invoked in assuming that (\ref{eq:thetainvar}) holds, i.e. that it does not make a difference to the photon traveling in the upper path whether the phase shifter is inserted or not. The assumption is akin to a kind of \emph{locality} constraint, as Hardy himself notes \cite[p.~3]{hardy2013}, i.e. informally: whether something is done in some region $A$ should not \emph{immediately} influence what happens in some none-overlapping region $B$. But of course we know that such an assumption becomes iffy in the context of QM, and hence it is doubtworthy whether any hidden variable model which purports to reproduce QM's predictions should respect it.

The critical reader will object that we have seen Spek-kens' toy model reproduce interferometer examples like the one considered in this section. Is the toy model non-local? \emph{Prima facie} the answer here is `no', but only on the price of accommodating a non-trivial `vacuum state', akin to that of quantum field theory (QFT) into the ontology presupposed by the model. Thus Hardy writes:\footnote{Here he is referring especially to elaborations from a talk given by Spekkens \cite{spekkens2008}.}

\begin{quote}[T]here are ontic variables associated with the occupation number of the path (take this to be 0 or 1) and a phase associated with the path (take this to be 0 or $\pi$). Even if the occupation number is 0 there is still the phase variable which will be affected by a phase shifter. Thus a path with no particle in it still has nontrivial degrees of freedom associated with it. This allows the model to violate ontic indifference in a local way. \cite[pp.~14-18]{hardy2013}\end{quote}

And Leifer similarly thinks that it is possible to save the interference examples from the consequences of Hardy's theorem in this fashion: 
\begin{quote}From quantum field theory, we know that the vacuum is not a featureless void, but has some sort of structure. Therefore, it makes sense that, at the ontological level, there might be more than one ontic state associated with the vacuum, and a transformation that does not affect things localized [in one arm of an interferometer -- FB] might still act nontrivially on these \emph{vacuum ontic states}. [...] A transformation acting locally on [one arm -- FB] can then switch the ontic states, in violation of ontic indifference, whilst leaving the distribution invariant. \cite[p.~121; my emphasis -- FB]{leifer2014}\end{quote}

So not: `something goes both ways', but rather: `for each way there is something which goes it'. But this analysis still has a foul taste to it and may strike us as somewhat in conflict with Spekkens' model. The reason is that here, for the first time, an appeal to the very nature of the true states is made, whereas the treatment so far has been entirely neutral on the subject. This may not be the biggest problem yet (indeed, we indicated that the model should at least allow for a specification of the nature of the true states), but the true states which are invoked as explanatory are `borrowed' from QM (more precisely: QFT), whereas so far QM was treated as precisely \emph{not} conveying a suitable ontology of the microcosm. Recall that Spekkens claims that: ``The key is that one can hope to identify phenomena that are characteristic of states of incomplete knowledge regardless of what this knowledge is about.'' \cite[p.~2]{spekkens2007} The interferometer example does obviously \emph{not} constitute an example of incomplete knowledge, \emph{regardless} of what this knowledge is about. For this example to make sense, one either has to commit to a direct influence between the two arms of the interferometer (which would violate the otherwise `local' spirit of the model), or one has to construct a specific kind of true state reminiscent of the vaccuum state from QFT, in order for the model to make sense. 

\section{Discussion: Prospects of an EE View of Quantum States}\label{sec:prospEinst}
What can we say about the plausibility of an EE view, in particular as presented in the form of $\psi$-epistemic OMs? To recall, the appeal of an EE view is that it is very natural and spares us a great deal of metaphysical complication. The appeal of the OM approach, on the other hand, is that it provides, in this context, a concrete formal framework to accommodate `Einsteinian' intuitions, and apparently allows to implement them in such a way that certain `problematic' predictions of QM are preserved or reproduced. 

Timpson, however, has objected that opting for hidden-variables in general
\begin{quote}is unlikely to be attractive to anyone who is trying to appeal to information as a way of avoiding the problems caused by the seemingly odd behaviour of the quantum state. The aim, roughly speaking, was to circumvent the problems associated with collapse or nonlocality by arguments of the form: there's not really any physical collapse, just a change in our knowledge; there's not really any nonlocality, it's only Alice's knowledge of (information about) Bob's system that changes when she performs a measurement on her half of an EPR pair. But we all know that if we are to have hidden variables lurking around then these are going to be very badly behaved indeed in quantum mechanics (nonlocality, contextuality). \cite[pp.~146-147; emphasis omitted]{timpson2013}\end{quote}

Of course the appeal here is to such theorems as that of Bell \cite{bell1964} or of Kochen and Specker \cite{kochenspecker1967}, and in the light of these one may ask: why even bother with hidden variables in the context of epistemic interpretations in the first place? In addition to these well known results we have here considered more specific no-gos aimed directly at $\psi$-epistemic models formulated in the OM approach. But we have argued that the impact of the influential PBR theorem may be compensated, and we have outlined some achievements of a $\psi$-epistemic toy model (and its more general spin off) in reproducing certain quantum phenomena and predictions from the mere assumption of epistemic restrictions. So what about these achievements?

As for quantum interference, we have seen that Spekkens can reproduce interference examples with his toy model, but that in the light of Hardy's theorem, one has to introduce specific elements into the model, namely what Leifer calls `vacuum ontic states', in order to produce an empirically adequate model without admitting non-locality. However, appealing to these vacuum ontic states may not be a recommendable move for the following reasons. 

First we must ask what a vacuum \emph{ontic} state actually \emph{is}. Of course this concept is supposed to incorporate the quantum field theoretic vacuum in the OM approach, but the vacuum in QFT is a \emph{theoretical concept}, i.e., what a `vaccum state' is ``cannot be fully specified by a single definition, but only by the joint effect of the core axioms of a \emph{theory}.'' \cite[p.~2; emphasis in original]{schurzgebharter2015} 

To make a case, consider the discussion of theoretical concepts by Schurz and Gebharter, who use the concept of \emph{force} in Newtonian physics as a paradigmatic example, which is defined, according to them, only by the joint axioms of Newtonian mechanics. A vacuum state is equally only defined by its role as the state of lowest eigenvalue for a given energy operator of some particular quantized field theory, and hence by the joint assumptions of the theory instead of one single theory-independent definition; and compare this also to \emph{individual} forces being defined in terms of particular differential equations and initial conditions for given physical problems. This dependence on a given field theory goes so far as to lead to two physically inequivalent vacui in the case of the Unruh effect \cite[cf.~e.g.][]{crispino2008}, for two observers who are non-inertially related to one another; and this may be compared to `ficticious forces' in Newtonian mechanics, which equally result from coordinate transformations between relatively non-inertial frames. It appears that no general, all-applying definition of the term `vacuum state' as used in QFT can be given, so that the situation is indeed comparable to that of Newtonian force. This should give some credibility for considering `vacuum state' as a theoretical concept of QFT in the aforementioned sense. 

Thus what a vacuum \emph{ontic} state \emph{is} is far from clear. It must be a \emph{new} concept, peculiar to a specific model, or rather: a suitable theory formulated by appeal to the OM approach, or a suitable replacement thereof. Without such a theory, the concept is not well-defined. But even on an intuitive level, the appeal seems completely \emph{ad hoc}, and one must wonder why it is introduced other than to facilitate certain philosophical preconceptions. 

What is (at least) equally bad, is that in algebraic QFT (AQFT), there is the \emph{Reeh-Schlieder theorem} \cite{reehschlieder1961}, which says that for an open bounded region $\mathcal{O}\subset\mathbb{R}^{4}$ of spacetime and $\hat{\mathrm{A}}(\mathcal{O})$ an element of the algebra generated from all possible combinations of adjoints, sums, and products of operators $\hat{\phi}(f) := \int\dd[4]{x}f(x)\hat{\phi}(x), f\in C^{\infty}_{0}(\mathcal{O})$, the set of vectors $\hat{\mathrm{A}}(\mathcal{O})\ket{\Omega}$ is dense in the space $\mathcal{H}$ of state vectors ($\ket{\Omega}$ the vacuum state). This means that one can approximate (arbitrarily close) \emph{any} state $\ket{\psi}$ by operations local to $\mathcal{O}$, even if $\ket{\psi}$ has implications for regions $\mathcal{O}'$ at a spacelike distance to $\mathcal{O}$ \cite[cf.][p.~S497~ff.]{fleming2000}. Most importantly, the theorem thus ``demonstrates'', as Dieks puts it, ``that the vacuum, and all other states of bounded energy, have long-distance correlations built into them. It is therefore not surprising to find that Bell inequalities are violated in these states---a standard sign of non-locality.'' \cite[p.~216]{dieks2002} 

Here Dieks of course refers to the work of Werner and Summers, who found, in the 1980s, ``that already the vacuum fluctuations assure a maximal violation of Bell's inequalities for the appropriate detectors.'' \cite[pp.~258-259]{wernersummers1985} Thus, any notion of `vacuum ontic states' that is sufficiently close to the QFT-notion of a vacuum state \emph{defects} the apparent locality of the interferometer examples---\emph{be-cause the element of QFT appealed to in order to restore locality is itself a decisive expression of quantum non-locality}.

One might object that these implications follow only from the highly theoretical algebraic version of QFT, and that in practice, the canonical quantization approach is all that is needed. This worry gains support by Wallace's observation that ``no examples are known of AQFT-compatible interacting field theories, and in particular the standard model cannot at present be made AQFT-compa-tible.'' \cite[p.~33]{wallace2006} Thus it may be suspected that these consequences of the Reeh-Schlieder theorem have no bearing on experimental practice and hence do not have to be taken seriously, in virtue of a lack of empirical accessibility. But similar worries were originally raised w.r.t.\! the strong non-local correlations predicted by ordinary QM (most notably by Schr\"odinger \cite[p.~166]{schrodinger1983}), and if there is anything we can learn from this example, it is that one is better off not to dismiss the implications of the quantum formalism easily. 

However, the situation is a bit more subtle in the case of testing vacuum entanglement, since, as Werner and Summers put it, ``there would be experimental difficulties [...][because] the violation of Bell's inequality must vanish exponentially with the spatial separation of [two separate spacetime regions] on the length scale determined by the Compton wavelength of the lightest particle of the theory.'' \cite[p.~259]{wernersummers1985} 

There are suggestions for other kinds of experiments in which vacuum states crucially enter into entangled states though, namely states entangled with those of a single photon. Examples of such experimental protocols are discussed, for instance, in \cite{tan1991} or \cite{hardy1994}. Typically these schemes are used to show that even a single particle is `nonlocal' in a sense, as noticed already by Einstein in his examples discussed at the 1927 Solvay conference \cite[cf.][pp.~115-116]{jammer1974}.

Despite some original controversy \cite[cf.][p.~2~ff. for discussion]{vedraldunnigham2007} today there is a broad consensus that particular experiments can be used to test exactly for this `single particle nonlocality' which involves entanglement with the vacuum, and the experiments that have been performed are reported to provide affirmations of the predictions \cite{hessmo2004, wiseman2015}. 

Provisios about the interpretation of the cited experiments aside, we hence have good theoretical \emph{and} empirical reasons to suspect that quantum vacuum states are just the kind of states which involve the problematic nonlocal correlations. In the light of these features of the QFT vacuum, we are lead to judge that the advocate of an EE view is faced with the following dilemma: if he appeals to vacuum states in close analogy to the vacuum state \emph{of QFT}, then he has \emph{neither} provided a local explanation of quantum interference phenomena after all, \emph{nor} (more importantly) followed his program of construing QM states as indicative of preparations and measurements only. If, on the other hand, he postulates a new kind of nontrivial vacuum, inspired by certain experimental results, he has merely shifted the burden from explaining interference to explaining these new kinds of states, \emph{together with all the empirical data that we have about QFT's vacuum states}. 

Let us assume, for the sake of argument, that the bullet is being bitten by taking the second horn of the dilemma and accepting a new kind of vacuum state, peculiar to a suitable $\psi$-epistemic OM (a thorough extension of Spekkens' model, say). Then we are lead to wonder: if such remarkable and remarkably counterintuitive results which can be \emph{derived} from quantum theory are simply \emph{presupposed} instead of explained and use-novelly predicted by a $\psi$-epistemic model, what good is the model then? 

Bartlett et al.\! in fact delineate their aim in seeking out epistemic models as follows: ``it is only by describing a broad landscape of possible theories that we can specify the sense in which quantum theory is special.'' \cite[p.~3]{bartlett2012} This is, indeed, an important task, and the OM approach in particular has already served to sort out some gos and no-gos for interpreting QM. But if this is the \emph{only} purpose of $\psi$-epistemic (and asorted) models, then they will \emph{not} help us solve (or resolve) the conceptual difficulties arising from the MP in QM by appeal to a \emph{mere} lack of knowledge about otherwise well-defined and independently existing entities and their states---in other words: they will not sketch a route to an interpretation that preserves the Einsteinian intuitions.  

Einstein viewed the QM of his time as ``no useful point of departure for future development'' \cite[p.~87]{schlipp1949}, and since the point of departure for the $\psi$-epistemic interpretations under consideration are Einsteinian worries, it must be seen as a `partial surrender' to QM if the only purpose of certain models is to show how QM is special. From an Einsteinian point of view, the aim must rather be to search for serious alternatives because QM is `too special'. 

One of Einstein's core intuitions clearly was locality or local causality (``Prinzip der Nahwirkung'' \cite[p.~322]{einstein1948}), which he held most dearly as a scientific principle (``A complete rejection of this principle would make the idea of the existence of (quasi-)closed systems and thereby the establishment of empirically testable laws in the sense familiar to us impossible.'' [ibid.; my translation -- FB]).\footnote{German original: ``V\"ollige Aufhebung dieses Grundsatzes w\"urde die Idee von der Existenz (quasi-)abgeschlossener Systeme und damit die Aufstellung empirisch pr\"ufbarer Gesetze in dem uns gel\"aufigen  Sinne unm\"oglich machen.''} So what about Spekkens' analysis of certain entangled states, which seemed to indicate that entanglement and correlations in remote measurement outcomes may also crucially involve a previous lack of knowledge about the true states of the systems involved? In fact,  Bartlett et al., in a similar vein, model ``maximal bipartite entanglement in [restricted Liouville] mechanics [...] by an epistemic state that describes perfect correlations between the pair of systems.'' \cite[p.~8]{bartlett2012} In particular, they use a probability distribution $\mu^{\mathrm{corr}}_{AB}(q_{A} ,p_{A} ,q_{B} ,p_{B}) \propto \delta(q_{A} -q_{B})\delta(p_{A} + p_{B})$ as phase space distribution for two systems $A$ and $B$ for which it is known that $q_{A}- q_{B} = 0$ and $p_{A} + p_{B} = 0$, i.e.\! which satisfy the conditions of the original EPR thought experiment \cite[cf.][]{epr1935}. Marginalizing for the coordinates of one of the two systems leads to a uniform distribution, so that nothing is known about the true states of the single systems, but only relational properties of the joint system are known (the total values for position and momentum). 

But this relational knowledge implies that in virtue of her prior knowledge of the value of the total momentum of the two systems, Alice, say, can determine the momentum value for Bob's system at once, after measuring momentum on her system, and analogously for position. In essence, we here get the same kind of informational update in consequence of a measurement on the total system, and thus just as the epistemic state in Spekkens' qubit-like toy model mirrored the properties of measurements on an entangled qubit state, the distribution in the restricted version of Liouville mechanics mirrors the properties of measurements on the original EPR-state.\footnote{\label{fn:EPRVIOL}There are, however, a few well known difficulties with the actual preparation and measurement of EPR states in the sense of the original paper: the state is not time dependent, and the descriptions used to set up the argument for incompleteness would only be valid at $t=0$, whereas time evolution makes it unstable; and since a plane wave representation is used, there would be non-vanishing probability of the two particles being basically anywhere in space, so that the assumption of spatial separatedness is actually unwarranted \cite[cf.][p.~13]{homeselleri1991}. However, Praxmeyer et al. have constructed a scheme in which the EPR state appears as the limit of a two-mode squeezed state, and observables on it are considered which can be used to violate a Bell-type inequality \cite{praxmeyer2005}.} And upon learning the value of Bob's position measurement (say), Alice can infer, on the basis of her own \emph{momentum} measurement, the complete phase space coordinates $(q_{A}, p_{A})$ at once, which makes EPR's original point.

In summary, as Bartlett et al. put it: 
\begin{quote}
All that changes as a result of this measurement is how the observer refines her knowledge of the ontic state of particle $B$. She either refines her knowledge of its position or she refines her knowledge of its momentum. No `spooky action at a distance' is required to understand the EPR experiment if one adopts the interpretation offered by [restricted Liouville] mechanics. \cite[p.~12]{bartlett2012}
\end{quote} 

Both examples, that of Spekkens and that of Bartlett et al., are indeed suggestive. But they are suggestive of something \emph{false}. It is \emph{not} that quantum non-locality can be explained in terms of knowledge \emph{in general}. It is only by \emph{selectively} choosing \emph{particular} states which can be mirrored by ordinary probability distributions that one can create the \emph{illusion} that this is possible. This is exactly the gist of Bell's theorem---that quantum \mbox{(anti-)}correlations are \emph{not} like those between Bertlmann's socks \cite{bell1981}. Indeed, \emph{admittedly} none of the models can reproduce violations of Bell-type inequalities or the like: 

\begin{quote}
The toy theory is, by construction, a local and noncontextual hidden variable theory. Thus, it cannot possibly capture all of quantum theory. In the face of these no-go theorems, a proponent of the epistemic view is forced to accept alternative possibilities for the nature of the ontic states to which our knowledge pertains in quantum theory. \citep[pp.~24-25]{spekkens2007}
\end{quote}

\begin{quote}
We emphasize that we are not arguing that a $\psi$-epistemic local hidden variable model could explain all quantum correlations, only that the particular correlations described in the EPR experiment can be so explained (in precisely the way that EPR suggested they should). This is not at odds with Bell's theorem because the correlations in the EPR experiment do not violate a Bell inequality.\footnote{Depending on the specific setup used to implement the states appealed to in the EPR paper, this becomes a debatable claim; cf. footnote~\ref{fn:EPRVIOL}.} Of course, because it is locally causal by construction, [restricted Liouville] mechanics cannot hope to reproduce Bell-inequality violations. Such violations are one of the quantum phenomena that [restricted Liouville] mechanics emphatically cannot reproduce, not even qualitatively. \citep[pp.~24-25]{bartlett2012}
\end{quote}

It seems clear that at the very least \emph{this} core intuition of Einstein (local causality) must be dropped in any empirically adequate $\psi$-epistemic model.  

\section{Conclusions}
In this paper we have analyzed a recent and influential approach that outlines paths to an epistemic reading of QM, and we have scrutinized its capability to provide specific models ($\psi$-epistemic models) that are compatible with `Einsteinian' intuitions. To this end, the one formal model (Spekkens' toy model) that so far has brought about the clearest conceptual successes in this direction was investigated in some detail, and a spin off for continuous degrees of freedom in all brevity. Subsequently, two recent no-go theorems were analyzed for their impact on the whole project.

We have found that the influential PBR theorem poses less of a threat than may be believed by some, but that it has spawned off an important debate and a plethora of further no-go theorems. We found Hardy's theorem, in particular, to pose a threat to certain achievements of Spekkens' toy model, a model that is sometimes considered to provide a \emph{plausibilization} of an EE view. In particular certain apparent successes of this model were demonstrated to rest on a selective choice of examples and/or to involve \emph{ad hoc} moves to such curiosities as `vacuum ontic states'. We have argued that (i) `vacuum ontic state' is effectively a theoretical concept without a proper theory, that (ii) such states may involve problematic features of non-locality (thus forestalling a local explanation of quantum interference in terms of hidden variables, as Spekkens' model is believed to deliver \cite{hardy2013, leifer2014}), and that (iii) an appeal to them should be considered illegitimate in virtue of (i), if \emph{quantum} states (including vacuum states $\ket*{\Omega}$) are \emph{merely} P/M-states. Given that Spekkens' model achieves a plausibilization of an EE view, these and similar arguments should count as a \emph{deplausibilization}.  

Moreover, we have found that the \emph{prima facie} explanations of specific EPR/Bell-examples in terms of incomplete knowledge were problematically selective, and that they eschewed the lesson of Bertlmann's socks. What we can see from all of this is that any meaningful OM is \emph{bound} to look like QM itself in important respects, and that if one strives for an ontology in addition to the instrumental/operational content of QM, then the quantum state is probably not `all and only epistemic'. All the creativity and formal elegance used in the general approach and the particular models discussed above apparently cannot bring us past this point. Indeed, more recently also Spekkens, holding firm to an epistemic view of the quantum state, has conceded that 

\begin{quote}the investigation of [epistemically restricted] theories is best considered as a first step in a larger research program wherein the framework of ontological models [...] is ultimately rejected, but where one holds fast to the notion that a quantum state is epistemic. \cite[p.~7]{spekkens2014}\end{quote}

We have here focused on the difficulties of an EE view of quantum states in spite of plausibilization strategies, and there are of course also aspects of QM that Spekkens' model and its continuous successor can handle much better than those focused on here \cite[cf.~in particular][for discussion]{jennings2015}. But quantum interference (being the method of choice to test for coherent superposition) and non-local correlations are arguably the most important quantum phenomena, and if these are not reproduced or only in a very unsatisfactory and \emph{ad hoc} fashion, it is not clear that the models discussed can serve as evidence for an EE view after all. Additionally, we once more emphasize that \emph{conceptually meaningful} models which also \emph{fully} reproduce QM are missing entirely to date, Spekkens' toy model and its successor for continuous degrees of freedom being maybe the closest calls. But notably even these two models do not at all explain what the true states $\lambda$ \emph{are}, or \emph{how} they bring about the puzzling quantum statistics and correlations; they merely stipulate their existence. The \emph{formally} successful models, on the other hand, in particular that of Lewis et al. \cite{lewisjennings2012}, appear, in philosophers' terms, \emph{gerrymandered}.

We must conclude at this point that the MP (to date) cannot be solved (or avoided) in the way we have called `natural', i.e.\! by depriving quantum states of their `onticity' and seeking for a deeper description in terms of hidden variables, more in accord with classical intuitions. This point remains throughout the decades, and it is due to a conflict with confirmed empirical predictions of QM. If the problems associated with QM are a matter of knowledge in some sense, then this knowledge still is knowledge about something rather peculiar. Thus if one is to have an epistemic interpretation of QM, one will either (a) have to accommodate quite a few features of QM into the underlying ontology, or (b) look to the other camp of epistemic interpretations, the BE ones (as advocated, e.g., in \cite{fuchs2014}).

An interesting route to (a) is hinted at in \cite{harriganspekkens2010}. Harrigan and Spek-kens refer to the possibility of reading Bohmian mechanics in the sense that the quantum state, $\psi$, of any subsystem of the universe is taken to be epistemic, whereas the quantum state of the whole universe, $\Psi$, is `nomological', as is advocated in particular by D\"urr et al. \cite[pp.~266~ff.]{durr2012}. The main reason for this distinction is that $\psi$ is subject to change, whereas $\Psi$ need not be (if it is a solution to the Wheeler-DeWitt equation, say; \cite[cf.][pp.~268-269]{durr2012}), and laws of nature supposedly should not change. But this approach to (a) seems problematic as well. If $\Psi$ describes a law of nature, then it is either part of an ontology (in all realistic and non-Humean views) and hence at least the quantum state of the universe is \emph{non-epistemic}, or, on a Humean such as that of Lewis, say, it should be a regularity appearing in a `best system', a system that ``strikes as good a balance as truth will allow between simplicity and strength'' \cite[p.~478]{lewis1994}, and therefor \emph{could} be interpreted epistemically as well. Since $\Psi$ is then a mere regularity though, this would, among other things, imply that particles simply correlate their behavior in certain experiments---\emph{they just do}.\footnote{Note also that if, on the other hand, one swallows the bitter pill and allows the hidden true states $\lambda$ to be `non-local' in a sense, then this might allow for superluminal signals becoming possible, in virtue of an envisioned future theory that includes more concrete descriptions of the $\lambda$s, and such signals are well-known to raise worries about causal paradoxes \cite[cf.][for discussion]{maudlin2011}.} More importantly, since there is no such best system to date in which $\Psi$ appears, it would be \emph{no one's} epistemic state. In any case, a good story needs to be told here as well. 

As a final remark to option (b), we note that in BE approaches the assumptions of (selective) scientific and metaphysical realism become iffy, because if we cannot even entertain dummy descriptions $\lambda$ to refer to the true configurations of that which is investigated in experiments, it indeed becomes at the very least \emph{unspeakable} \cite{bell1987}. Put frankly, if the quantum state is not the true state of the system, and there is also no additional true state $\lambda$, then \emph{maybe there is no true state of the system}. This is not only in defiance of the semantic condition of scientific realism, but it also raises doubts about the very \emph{existence} (in a mind-, \mbox{language-,} and theory-independent sense) of fundamental physical entities, and thus about metaphysical realism. 

As regards the main subject of this paper, we may sum up the worries at this point with a mutilated version of one of Einstein's own comments on Schr\"odinger's wave mechanics: 

\begin{quote}The successes of [Harrigan and Spekkens'] theory make a great impression, and yet [we] do not know whether it is question [\emph{sic}] of anything more than the old quantum rules [...]. Has one really come closer to a solution of the riddle? (after Einstein 1926; as cited in \cite[pp.~83-84]{howard1990})\end{quote}

Maybe Bohr will win this one too... 
 \section*{Acknowledgement} The research was financially supported by the Philosophy Department, University of D\"usseldorf, Prof. Markus Schrenk. I am also grateful to two anonymous commentators for helpful commentary.
  \section*{References} 
\bibliography{QEpist}

\end{document}